\documentclass[prd,twocolumn,tightenlines,showpacs,nofootinbib]{revtex4-1} 

\usepackage{amsfonts}
\usepackage{dcolumn}
\usepackage{bm}
\usepackage{amsmath}
\usepackage{nicefrac}
\usepackage{amsthm}
\usepackage{amssymb}
\usepackage{graphicx}
\usepackage{color}

\newcommand{\appropto}{\mathrel{\vcenter{
  \offinterlineskip\halign{\hfil$##$\cr
    \propto\cr\noalign{\kern2pt}\sim\cr\noalign{\kern-2pt}}}}}

\begin{document}

\title{Searching for dark matter and variation of fundamental constants with laser and maser interferometry}

\date{\today}
\author{Y.~V.~Stadnik} \email[]{y.stadnik@unsw.edu.au}
\affiliation{School of Physics, University of New South Wales, Sydney 2052,
Australia}
\author{V.~V.~Flambaum} \email[]{v.flambaum@unsw.edu.au}
\affiliation{School of Physics, University of New South Wales, Sydney 2052,
Australia}

\begin{abstract}
Any slight variations in the fundamental constants of Nature, which may be induced by dark matter or some yet-to-be-discovered cosmic field, would characteristically alter the phase of a light beam inside an interferometer, which can be measured extremely precisely. Laser and maser interferometry may be applied to searches for the linear-in-time drift of the fundamental constants, detection of topological defect dark matter through transient-in-time effects and for a relic, coherently oscillating condensate, which consists of scalar dark matter fields, through oscillating effects. Our proposed experiments require either minor or no modifications of existing apparatus, and offer extensive reach into important and unconstrained spaces of physical parameters.
\end{abstract}


\maketitle 


The idea that the fundamental constants of Nature might vary with time can be traced as far back as the large numbers hypothesis of Dirac, who hypothesised that the gravitational constant $G$ might be proportional to the reciprocal of the age of the Universe \cite{Dirac1937}. More contemporary theories predict that the fundamental constants vary on cosmological timescales (see e.g.~\cite{Terazawa1981,Uzan2002,Calmet2002}). Astronomical observations of quasar absorption spectra hint at the existence of a spatial gradient in the value of the fine-structure constant, $\alpha = e^2 / \hbar c$ \cite{Webb2011,King2012}. Data samples from the  Keck Telescope and Very Large Telescope  \cite{Webb1999,Webb2003}
 independently agree on the direction and magnitude of this gradient, which is significant at the $4.2 \sigma$ level. A consequence of this astronomical result is that, since the solar system is moving along this spatial gradient, there should exist a corresponding temporal shift in $\alpha$ in Earth's frame of reference at the level $\delta \alpha/\alpha \sim 10^{-19}$/yr \cite{Berengut}. Finding this variation with laboratory experiments could independently corroborate the astronomical result. To date, atomic clocks have provided the most sensitive laboratory limit on annual variations in $\alpha$: $\delta \alpha/\alpha \lesssim 10^{-17}$/yr \cite{Rosenband2008}.

The question of dark matter (DM), namely its identity, properties and non-gravitational interactions, remains one of the most important unsolved problems in physics. Various DM candidates and searches therefor have been proposed over the years \cite{Bertone2010Book}. One such candidate is the axion, a pseudoscalar particle which was originally introduced in order to resolve the strong CP problem of Quantum Chromodynamics (QCD) \cite{Peccei1977,Peccei1977b} (see also \cite{Kim1979,Shifman1980,Zhitnitsky1980,Dine1981}). The axion is believed to have formed a condensate in the early Universe \cite{Sikivie2009}. This relic axion condensate can be sought for through a variety of distinctive signatures (see e.g.~\cite{Sikivie1983,ADMX2010,Graham2011,Graham2013,Stadnik2014axions,CASPER2014,Roberts2014prl,Roberts2014long}). Likewise, a condensate consisting of a scalar DM particle may also have formed. The scalar field $\eta$ comprising this condensate oscillates with frequency $\omega \approx m_\eta c^2 / \hbar$ and may couple to the fermion fields:
\begin{equation}
\label{scalar-fermion_osc}
\mathcal{L}_{\textrm{int}}^{f} = - \sum_{f=e,p,n} \eta_0 \cos(m_\eta c^2 t/ \hbar) \frac{ m_f c^2 }{\Lambda_f} \bar{f}f ,
\end{equation}
where $f$ is the fermion Dirac field and $\bar{f}=f^\dagger \gamma^0$, and to the electromagnetic field:
\begin{equation}
\label{scalar-photon_osc}
\mathcal{L}_{\textrm{int}}^{\gamma} =   \frac{\eta_0 \cos(m_\eta c^2 t/ \hbar)}{4 \Lambda_\gamma} F_{\mu \nu} F^{\mu \nu} ,
\end{equation}
where $F$ is the electromagnetic field tensor. $\Lambda_X$ is a large energy scale, which from gravitational tests is constrained to be $\Lambda_X \geq 10^{21}$ GeV \cite{Derevianko2014}. Eqs.~(\ref{scalar-fermion_osc}) and (\ref{scalar-photon_osc}) alter the fundamental constants in an oscillating manner as follows, respectively:
\begin{equation}
\label{delta-m_f}
m_f \to m_f \left[1+ \frac{\eta_0 \cos(m_\eta c^2 t/ \hbar)}{\Lambda_f} \right] ,
\end{equation}
\begin{equation}
\label{delta-alpha}
\alpha \to \frac{\alpha}{1 - \eta_0 \cos(m_\eta c^2 t/ \hbar) / \Lambda_\gamma } \simeq \alpha \left[1+ \frac{\eta_0 \cos(m_\eta c^2 t/ \hbar)}{\Lambda_\gamma} \right] .
\end{equation}
Quadratic couplings in $\eta$, with the replacement $\eta_0 \cos(m_\eta c^2 t/ \hbar) / \Lambda_X \to [\eta_0 \cos(m_\eta c^2 t/ \hbar) / \Lambda'_X ]^2$ in Eqs.~(\ref{scalar-fermion_osc}) -- (\ref{delta-alpha}), are also possible. $\Lambda_X '$ is a less strongly constrained energy scale, with constraints from laboratory and astrophysical observations given by $\Lambda_X ' \geq 10^4$ GeV \cite{Olive2008}. We note that the quadratic portal gives rise to not only oscillating effects, but may also lead to non-oscillating space-time variation of the fundamental constants: $\delta X/ X = \eta_0^2 / 2 (\Lambda'_X)^2$, which arises due to space-time variations in $\left<\eta^2 \right>$. These effects may be sought for using terrestrial experiments (atomic clocks, Oklo natural nuclear reactor, and laser/maser interferometry as suggested in this paper) and astrophysical observations (quasars, white dwarves, Big Bang Nucleosynthesis, Cosmic Microwave Background measurements).

Another possible DM candidate is topological defect DM, which is a stable non-trivial form of DM that consists of light DM fields and is stabilised by a self-interaction potential \cite{Vilenkin1985} (self-gravitating DM fields can also form solitons, see e.g.~Ref.~\cite{Marsh2015soliton}). These objects may have various dimensionalities: 0D (monopoles), 1D (strings) or 2D (domain walls). The transverse size of a topological defect depends on the mass of the particle comprising the defect, $d \sim \hbar/m_\phi c$, which may be large (macroscopic or galactic) for a sufficiently light DM particle. The light DM particle comprising a topological defect can be either a scalar, pseudoscalar or vector particle. Recent proposals for pseudoscalar-type defect searches include using a global network of magnetometers to search for correlated transient spin precession effects \cite{GNOME2013} and electric dipole moments \cite{Stadnik2014defects} that arise from the coupling of the scalar field derivative to the fermion axial vector currents. Recent proposals for scalar-type defect searches include using a global network of atomic clocks \cite{Derevianko2014}, and Earth rotation and pulsar timing \cite{Stadnik2014defects}, to search for transient-in-time alterations of the system frequencies due to transient-in-time variation of the fundamental constants that arise from the couplings of the scalar field to the fermion and photon fields. The best current sensitivities for transient-in-time variations of the fundamental constants on the time scale of $t \sim 1 - 100$ s with terrestrial experiments are offered by atomic clocks, with an optical/optical clock combination \cite{Hinkley2013,Bloom2014} sensitive to variations in $\alpha$: $\delta \alpha / \alpha \sim 10^{-15} - 10^{-16}$ and a hyperfine/optical clock combination \cite{Jefferts2013} to variations in the electron-to-proton mass ratio $m_e/m_p$: $\delta (m_e/m_p) / (m_e/m_p) \sim 10^{-13} - 10^{-14}$. 

There are many possibilities for the interactions of topological defect DM particles with the Standard Model particles. Here we consider couplings with a quadratic dependence on the scalar field, which were considered previously in Refs.~\cite{Stadnik2014defects,Derevianko2014}. A scalar dark matter field $\phi$ may interact with fermions via the coupling:
\begin{equation}
\label{scalar-fermion_TD}
\mathcal{L}_{\textrm{int}}^{f} = - \sum_{f=e,p,n} m_f \left(\frac{\phi c}{\Lambda_f'}\right)^2 \bar{f} f ,
\end{equation}
and with photons via the coupling:
\begin{equation}
\label{scalar-photon_TD}
\mathcal{L}_{\textrm{int}}^{\gamma} = \left(\frac{\phi}{\Lambda_\gamma'}\right)^2 \frac{F_{\mu \nu} F^{\mu \nu}}{4} ,
\end{equation}
Eqs.~(\ref{scalar-fermion_TD}) and (\ref{scalar-photon_TD}) alter the fundamental constants in a transient manner as follows, respectively: 
\begin{equation}
\label{delta-m_f_TD}
m_f \to m_f \left[1+ \left(\frac{\phi}{\Lambda_f '}\right)^2 \right] ,
\end{equation}
\begin{equation}
\label{delta-alpha_TD}
\alpha \to \frac{\alpha}{1 - (\phi/\Lambda_\gamma ')^2 } \simeq \alpha \left[1+ \left(\frac{\phi}{\Lambda_\gamma '}\right)^2 \right] .
\end{equation}

In the present work, we point out that laser and maser interferometry may be used as particularly sensitive probes to search for linear-in-time, oscillating and transient variations of the fundamental constants of Nature, including $\alpha$ and $m_e/m_p$. Laser and maser interferometry are very well established techniques and have already proven to be extremely sensitive probes for exotic new physics, including searches for the aether, tests of Lorentz symmetry \cite{Kostelecky2011RMP} and gravitational wave detection \cite{LIGOBook2014}. Laser interferometry has also recently been proposed for the detection of dilaton dark matter \cite{VanTilburg2015}. 

We consider the use of an interferometer with two arms of lengths $L_1$ and $L_2$, for which the observable is the phase difference $\Delta \Phi = \omega \Delta L / c$ between the two split beams, where $\omega$ is the reference frequency and $\Delta L = L_1 - L_2$. In the absence of any variation of fundamental constants, the two split beams interfere destructively ($\Delta \Phi = (2N+1)\pi$, where $N$ is an integer). In the presence of variation of the fundamental constants, the reference frequency changes, as do the arm lengths, due to changes in the sizes of the atoms, which make up the arms. Depending on the type of laser or maser, as well as the arm lengths and materials used, the net result may be a change in the phase difference, $\delta (\Delta \Phi)$.

Consider the simpler case when a laser/maser without a resonator is used, for example, the nitrogen laser operating on the $^3 \Pi_u ~\to~ ^3\Pi_g$ electronic transition and superradiant Raman lasers \cite{Holland2009,Thomson2012a,Thomson2012b}. In this case, $\omega$ is determined entirely by the specific atomic/molecular transition, the simplest archetypes of which are the electronic Rydberg ($\omega \sim e^2/a_B \hbar$), hyperfine ($\omega \sim (e^2/a_B \hbar)( m_e/m_p)  \alpha^{2+K_{\textrm{rel}}} \mu$), vibrational ($\omega \sim (e^2/a_B \hbar)\sqrt{m_e/m_p M_r}$) and rotational ($\omega \sim (e^2/a_B \hbar)(m_e/m_p M_r)$) transitions, where $\mu$ is the relevant nuclear magnetic dipole moment, $K_{\textrm{rel}}$ is the derivative of the hyperfine relativistic (Casimir) correction factor with respect to $\alpha$, and $m_p M_r$ is the relevant reduced mass. The sensitivity coefficients $K_X$ are defined by
\begin{equation}
\label{sens_coeffs}
\frac{\delta (\Delta \Phi)}{\Delta \Phi} = \sum_{X=\alpha,m_e/m_p,m_q/\Lambda_{\textrm{QCD}}} K_X \frac{\delta X}{X} ,
\end{equation}
where $m_q$ is the quark mass and $\Lambda_{\textrm{QCD}}$ is the QCD scale, and are given in Table \ref{tab:sens-coeffs} for several archetypal transitions, where we have made use of the relation $\delta(\Delta L) / \Delta L \approx \delta a_B / a_B$ for $\Delta L \ne 0$ ($a_B$ is the Bohr radius).
Since $\delta (\Delta \Phi)$ is proportional to $\Delta \Phi$, a higher laser frequency gives a larger effect.

  \begin{table}[h!]
    \centering%
    \caption{Sensitivity coefficients for $\alpha$, $m_e/m_p$ and $m_q/\Lambda_{\textrm{QCD}}$ for a laser or maser without a resonator and operating on typical atomic and molecular transitions. The values of $K_{\textrm{rel}}$ and $K_{m_q/\Lambda_{\textrm{QCD}}}$ for $^{87}$Rb and $^{133}$Cs have been taken from \cite{Tedesco2006} (see also Refs.~\cite{Flambaum1999Theor,Dinh2009}).} 
\begin{ruledtabular}%
\begin{tabular}{cccc}
Transition & $K_\alpha$ & $K_{m_e/m_p}$ & $K_{m_q/\Lambda_{\textrm{QCD}}}$ \\
\hline
Electronic & 1 & 0 & 0 \\
Hyperfine ($^{87}$Rb) & 3.34 & 1 & -0.016 \\
Hyperfine ($^{133}$Cs) & 3.83 & 1 & 0.009 \\
Vibrational & 1 & 1/2 & 0 \\
Rotational & 1 & 1 & 0 \\  
  \end{tabular}%
\end{ruledtabular}%
    \label{tab:sens-coeffs}%
  \end{table}

Note that, unlike atomic clock experiments \cite{Rosenband2008,Gill2014,Peik2014} and astrophysical observations \cite{Webb1999,Ubachs2006,Webb2011} that search for a variation in the fundamental constants, in which two different transition lines are required to form the dimensionless ratio $\omega_A / \omega_B$, laser/maser interferometry can in principle be performed with only a single line, since the observable $\Delta \Phi$ is a dimensionless parameter by itself. However, one may also perform two simultaneous interferometry experiments with two different transition lines, using the same set of mirrors. Treating variations in frequencies (which depend only on the fundamental constants) and lengths independently (for variations in the latter may also arise due to undesired effects), we find
\begin{equation}
\label{delta_X-big}
\delta X = \frac{c \left[\omega_A \delta (\Delta \Phi_B) - \omega_B \delta(\Delta \Phi_A) \right]}{\Delta L \left( \omega_A \frac{\partial \omega_B}{\partial X} - \omega_B \frac{\partial \omega_A}{\partial X}  \right)} ,
\end{equation}
where $X$ is a particular fundamental constant. In particular, we note that shifts in the arm lengths do not appear in Eq.~(\ref{delta_X-big}), meaning that undesirable effects, such as seismic noise or tidal effects, are not observed with this setup and high precision may in principle be attained for low-frequency (large timescale) effects. This is quite distinct from conventional interferometer searches for gravitational waves, which have comparatively low sensitivity to low-frequency effects, since in this case deviations in arm lengths are sought explicitly and low-frequency systematic effects greatly reduce the sensitivity of the apparatus in this region.


Consider now the case when a laser/maser containing a resonator is used, for instance, the Nd:YAG solid-state laser. In this case, $\omega$ is determined by the length of the resonator, which changes if the fundamental constants change. In the non-relativistic limit, the wavelength and $\Delta L$ (as well as the size of Earth) have the same dependence on the Bohr radius and so there are no observable  effects if changes of the fundamental constants are slow (adiabatic).  Indeed, this may be viewed as a simple change in the measurement units. Transient effects due to topological defect DM passage may still produce effects, since changes in $\omega$ and $\Delta L$ may occur at different times. We note that a global terrestrial network (LIGO, Virgo, GEO600 and TAMA300) or a space-based network of interferometers (LISA) are particularly well suited to search for topological defects through the correlated effects induced by defects. Likewise, temporal correlations of homogeneous effects (including linear-in-time and oscillating effects) produced in several different interferomers can also be sought for.

The sensitivity of interferometry to non-transient effects is determined by relativistic corrections, which we estimate as follows. The size of an atom $R$ is determined by the classical turning point of an external atomic electron. Assuming that the centrifugal term $\sim 1/R^2$ is small at large distances, we obtain  $(Z_i+1)e^2/R = |E|$, where $E$ is the energy of the external  electron and $Z_i $ is the net charge of the atomic species (for a neutral atom $Z_i=0$). This gives the relation: $\delta R/ R = - \delta |E| / |E|$. The single-particle relativistic correction to the energy in a many-electron atomic species is given by \cite{Flambaum1999}:
\begin{equation}
\label{rel-correxn_FS}
\Delta_n \simeq E_n \frac{(Z\alpha)^2}{\nu (j+1/2)} ,
\end{equation}
where $E_n = - m_e e^4 (Z_i+1)^2 / 2 \hbar^2 \nu^2$ is the energy of the external atomic  electron, $j$ is its angular momentum, $Z$ is the nuclear charge,  and $\nu \sim 1$ is the effective principal quantum number. The corresponding sensitivity coefficient in this case is
\begin{equation}
\label{sens_coeffn_alpha_complex}
K_\alpha = 2\alpha^2 \left[\frac{Z_{\textrm{res}}^2}{\nu_{\textrm{res}} (j_{\textrm{res}} + 1/2)} - \frac{Z_{\textrm{arm}}^2}{\nu_{\textrm{arm}} (j_{\textrm{arm}} + 1/2)} \right]  . 
\end{equation}
Note that the sensitivity coefficient depends particularly strongly on the factor $Z^2$. $\left|K_\alpha\right| \ll 1$ for light atoms and may be of the order of unity in heavy atoms. Furthermore, the arms of different length can also be replaced by two arms (of the same length) made from different materials, for which the coefficients $Z^2/\nu(j+1/2)$ are different. 

We estimate the sensitivity to variations in $m_e/m_p$ from the differences in the internuclear separations in molecular H$_2$ and D$_2$, which are $0.74144 ~ \textrm{\AA}$ and $0.74152 ~ \textrm{\AA}$, respectively \cite{NIST-database}. These data give: $\delta R/ R \approx - 10^{-4} ~ \delta(m_e/m_p) / (m_e/m_p)$. Since only differences in the coefficients of proportionality for the arm and resonator are observable in principle, the corresponding sensitivity coefficient is therefore $\left|K_{m_e/m_p} \right| \lesssim 10^{-4}$. 

Note that for a slow variation of fundamental constants (which includes linear-in-time effects, transient effects due to a slowly moving and/or large topological defect, and low-frequency oscillating effects), the laser/maser resonator may be locked to an atomic/molecular frequency. In these cases, the sensitivity coefficients will be the same as those for the case in which a laser/maser without a resonator is used.

\begin{figure}[h!]
\begin{center}
\includegraphics[width=8.5cm]{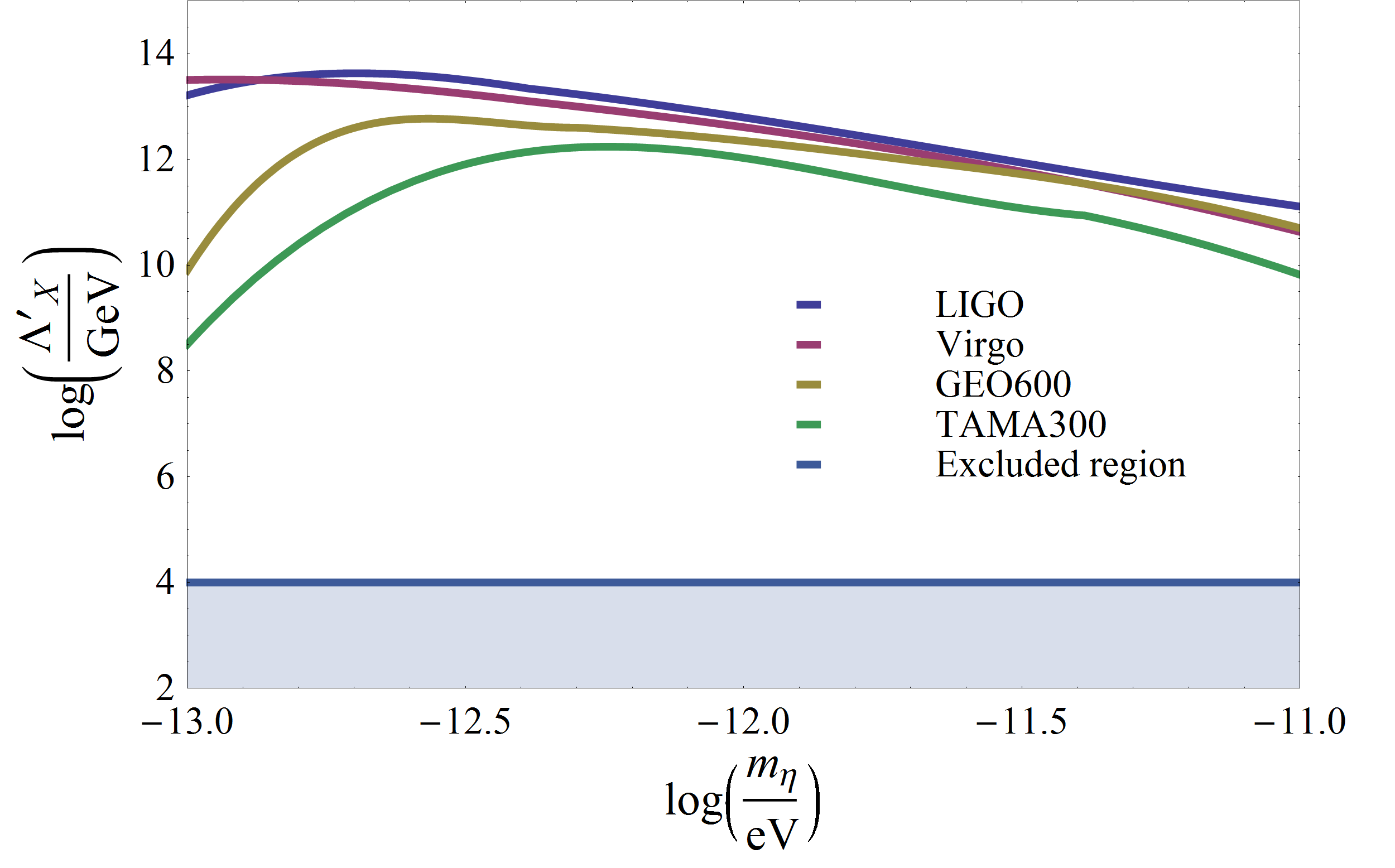}
\caption{(Color online) Region of dark matter parameter space accessible by various interferometers. The shaded blue region corresponds to the region of parameter space excluded by existing laboratory and astrophysical observations \cite{Olive2008}.} 
\label{fig:condensate_sens}
\end{center}
\end{figure}

We estimate the sensitivity of laser and maser interferometry to effects stemming from a relic, coherently oscillating condensate, which consists of scalar DM fields. The typical spread in the oscillation frequencies of the scalar DM particles, which make up the condensate, is given by $\Delta \omega / \omega \sim (\frac{1}{2} m_\eta v^2 / m_\eta c^2) \sim (v^2/c^2)$, where a virial velocity of $v \sim 10^{-3} c$ would be typical in our local Galactic neighbourhood. From the strain sensitivity curves of various interferometers \cite{LIGO2012,GEO2014,TAMA2004}, and assuming that the condensate consisting of a scalar DM particle saturates the known local cold DM content, $\eta_0^2 m_\eta^2 c^2 / 2 \hbar^2 \sim 0.4$ GeV/cm$^{3}$, we arrive at the accessible region of parameter space shown in Fig.~\ref{fig:condensate_sens}, in which we assume the use of a laser without a resonator. The region of parameter space accessible by the recently constructed Fermilab Holometer ($L=40$ m) \cite{FHolometer} is expected to be similar to those accessible by the interferometers as shown in Fig.~\ref{fig:condensate_sens}, but shifted toward higher scalar DM masses by several orders of magnitude.

\begin{figure}[h!]
\begin{center}
\includegraphics[width=8.5cm]{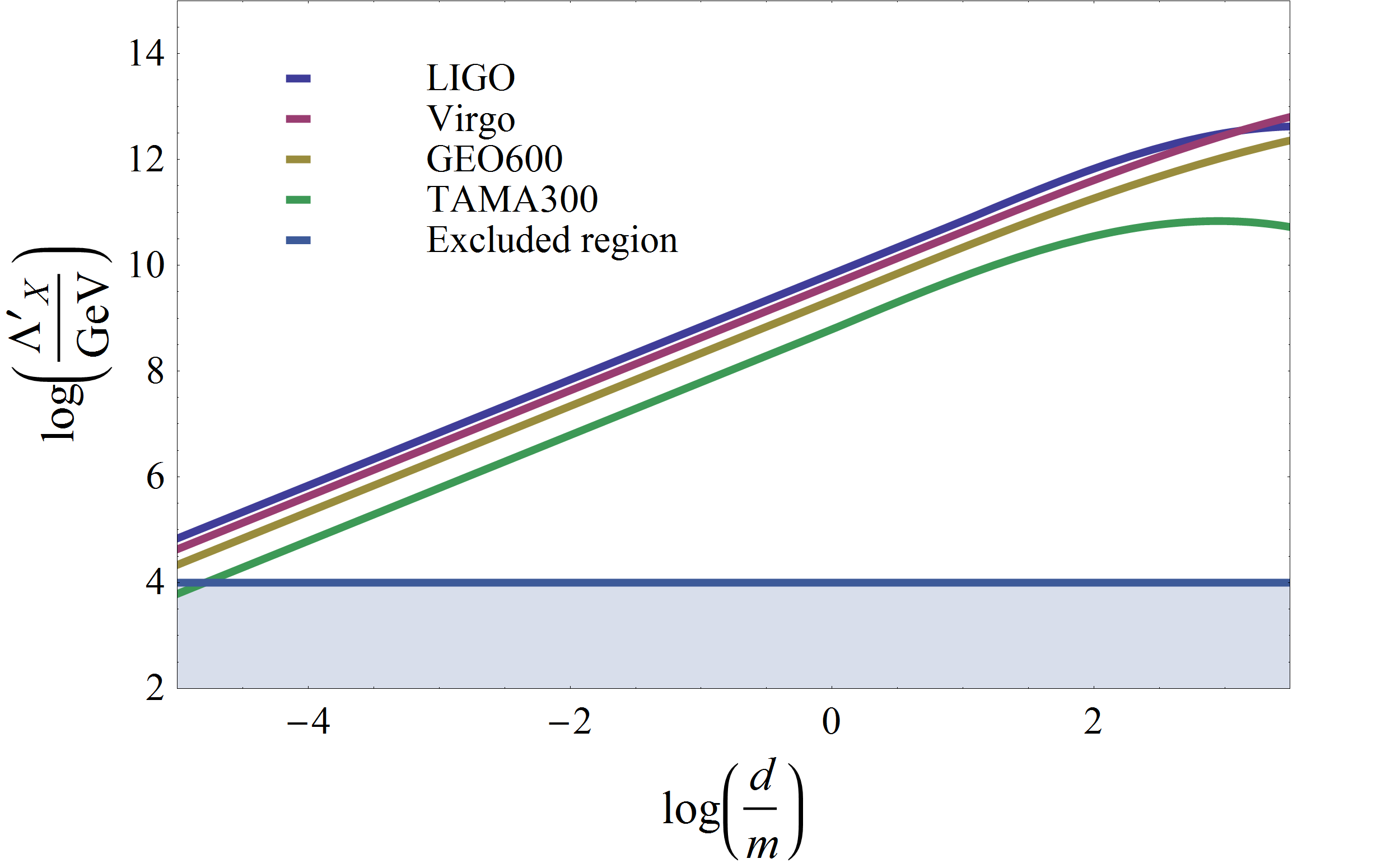}
\caption{(Color online) Region of dark matter parameter space accessible by various interferometers. The shaded blue region corresponds to the region of parameter space excluded by existing laboratory and astrophysical observations \cite{Olive2008}.} 
\label{fig:defect_sens}
\end{center}
\end{figure}

Finally, we estimate the sensitivity of laser and maser interferometry to effects stemming from topological defects, which consist of scalar DM fields, using the simple model of a domain wall with a Gaussian cross-sectional profile of root-mean-square width $d$. The simplest domain wall direction of incidence to consider (which produces non-zero effects) is directly along one of the interferometer arms (towards the laser source, without loss of generality). Neglecting relativistic effects and assuming the use of a resonator-based laser, the time-domain signal is given by
\begin{align}
\label{Time-domain_signal}
&\frac{\delta [\Delta \Phi (t)]}{\Delta \Phi} = \frac{\rho_{\textrm{TDM}} v \tau d \hbar c}{(\Lambda_X')^2} \cdot \notag \\
&\left\{ \frac{d\sqrt{\pi}}{2L} \left[\textrm{erf}\left(\frac{L+tv}{d} \right) - \textrm{erf}\left(\frac{tv}{d} \right) \right] - \textrm{exp} \left(\frac{-t^2 v^2}{d^2} \right) \right\}  ,
\end{align}
where $\rho_{\textrm{TDM}}$ is the the energy density associated with a topological defect network, $v$ is the typical speed of a defect, $\tau$ is the average time between encounters of a system with defect objects, and erf is the standard error function. Cosmological models of topological defect DM have sufficient flexibility for topological defects to be the dominant contributor to the total DM content of the universe \cite{Derevianko2014}. For the purposes of estimating the sensitivity of laser interferometers to topological defects, we may hence assume $\rho_{\textrm{TDM}} \sim 0.4$ GeV/cm$^{3}$. Also, from hints offered by pulsar timing data in relation to the pulsar glitch phenomenon \cite{Stadnik2014defects}, we assume $\tau \sim 1$ year. The power spectrum corresponding to the time-domain signal in Eq.~(\ref{Time-domain_signal}) is given by
\begin{widetext}
\begin{align}
\label{power_spectrum_general}
&\left|\frac{\delta [\Delta \Phi (f)]}{\Delta \Phi}\right|^2 \approx \frac{\rho_{\textrm{TDM}}^2 v^2 \tau^2 d^4 \hbar^2 c^2}{16 \pi (\Lambda_X')^4} \cdot 
 \left|\frac{i e^{-\frac{\pi  f \left(\pi  d^2 f+4 i d v-2 i L  v\right)}{v^2}}}{f L}  \left[\text{erf}\left(\frac{L}{d}+2\right) e^{\frac{\pi  f \left(\pi  d^2 f-2 i L v\right)}{v^2}}-e^{\frac{4 i \pi  d f}{v}}  \text{erf}\left(\frac{i \pi  d f}{v}+\frac{L}{d}+2\right) \right. \right. \notag \\
&\left. + \text{erf}(2)  e^{\frac{\pi  d f (\pi  d f+8 i v)}{v^2}}+ e^{\frac{4 i \pi  d f}{v}} \text{erf}\left(\frac{i \pi  d f}{v}-2\right)\right]  -  \frac{2 \pi  e^{-\frac{\pi ^2 d^2 f^2}{v^2}}}{v} \left[\text{erf}\left(-\frac{i \pi  d  f}{v}+\frac{L}{d}+2\right)+\text{erf}\left(2+\frac{i \pi  d  f}{v}\right)\right]   \notag \\
&\left. - \frac{i e^{-\frac{\pi  d f (\pi  d f+4 i v)}{v^2}}}{f L} \left[\text{erf}(2) e^{\frac{\pi ^2 d^2  f^2}{v^2}}+\text{erf}\left(\frac{L}{d}+2\right) e^{\frac{\pi  f \left(\pi  d^2  f+8 i d v+2 i L v\right)}{v^2}}-e^{\frac{4 i \pi  d f}{v}} \text{erf}\left(-\frac{i \pi  d f}{v}+\frac{L}{d}+2\right)-e^{\frac{4 i \pi  d  f}{v}} \text{erf}\left(2+\frac{i \pi  d f}{v}\right)\right]\right| ^2
,
\end{align}
\end{widetext}
with the following asymptotic limit when $d \gg L$:
\begin{align}
\label{power_spectrum_large_d}
\left|\frac{\delta [\Delta \Phi (f)]}{\Delta \Phi}\right|^2 \sim \frac{\pi^3 \rho^2_{\textrm{TDM}} \tau^2 d^4  L^2 f^2 \hbar^2 c^2 \textrm{exp} \left(\frac{-2 \pi^2 f^2 d^2}{v^2}\right)}{(\Lambda_X')^4 v^2} .
\end{align}
From the power spectrum in (\ref{power_spectrum_general}), the plots for which are presented for interferometers of various sizes in the Supplemental Material, and the strain sensitivity curves of these interferometers \cite{LIGO2012,GEO2014,TAMA2004}, we arrive at the accessible region of parameter space shown in Fig.~\ref{fig:defect_sens}. We note that the sensitivity of interferometers drops rapidly with increasing values of $d$ when $d \gtrsim L$. For instance, for a LIGO interferometer ($L=4$ km), the sensitivity to defects with $d=40$ km is $\Lambda'_X \lesssim 10^{-4}$ GeV. The region of parameter space accessible by the Fermilab Holometer \cite{FHolometer} is expected to be similar to those accessible by the interferometers as shown in Fig.~\ref{fig:defect_sens}, but with a rapid drop in sensitivity occurring for $d \gtrsim 100$ m.

We hence suggest the use of laser and maser interferometry as particularly sensitive probes to search for linear-in-time, oscillating and transient variations of the fundamental constants of Nature, including $\alpha$ and $m_e/m_p$. Our proposed experiments require either minor or no modifications of existing apparatus, and offer extensive reach into important and unconstrained spaces of physical parameters. We note that oscillating variation of fundamental constants due to a scalar condensate may also be sought for using atomic clocks.

\textbf{Acknowledgements} --- We would like to thank Francois Bondu, Dmitry Budker, Sergey Klimenko and Guenakh Mitselmakher for helpful discussions. This work was supported by the Australian Research Council.




\begin{thebibliography}{99} 





\bibitem{Dirac1937} Dirac, P.~A.~M.~The Cosmological Constants. \emph{Nature (London)} \textbf{139,} 323-323 (1937).


\bibitem{Terazawa1981} H.~Terazawa, Cosmological Origin of Mass Scales. \emph{Phys.~Lett.~B} \textbf{101}, 43 (1981).

\bibitem{Uzan2002} Uzan, J.-P.~The fundamental constants and their variation: observational and theoretical status. \emph{Rev.~Mod.~Phys.~}\textbf{75,} 403 (2003).

\bibitem{Calmet2002} Calmet, X. and Fritzsch, H.~The cosmological evolution of the nucleon mass and the electroweak coupling constants. \emph{Eur. Phys. J. C} \textbf{24}, 639 (2002).

\bibitem{Webb2011} Webb, J.~K.~\emph{et.~al.~}Indications of a Spatial Variation of the Fine Structure Constant. \emph{Phys.~Rev.~Lett.}~\textbf{107,} 191101 (2011).

\bibitem{King2012} King, J.~A.~\emph{et.~al.~}Spatial variation in the fine-structure constant - new results from VLT/UVES. \emph{MNRAS} \textbf{422,} 3370 (2012).

\bibitem{Webb1999} Webb, J.~K., Flambaum, V.~V., Churchill, C.~W., Drinkwater, M.~J.~\& Barrow, J.~D.~Search for time variation of the fine structure constant. \emph{Phys.~Rev.~Lett.~}\textbf{82,} 884-887 (1999).

\bibitem{Webb2003} Murphy, M.~T., Webb, J.~K.~\& Flambaum, V.~V.~Further evidence for a variable fine structure constant from Keck/HIRES QSO absorption spectra. \emph{MNRAS} \textbf{345,} 609 (2003).

\bibitem{Berengut} Berengut, J.~C.~\& Flambaum, V.~V.~Manifestations of a spatial variation of fundamental constants in atomic and nuclear clocks, Oklo, meteorites, and cosmological phenomena. \emph{Europhys.~Lett.~}\textbf{97,} 20006 (2012).

\bibitem{Rosenband2008} Rosenband, T.~\emph{et.~al.~}Frequency Ratio of Al$^{+}$ and Hg$^+$ Single-Ion Optical Clocks; Metrology at the 17th Decimal Place. \emph{Science} \textbf{319,} 1808-1812 (2008).

\bibitem{Bertone2010Book} Bertone, G.~(Ed.) \emph{Particle Dark Matter: Observations, Models and Searches}. (Cambridge University Press, Cambridge, 2010).

\bibitem{Peccei1977} Peccei, R.~D.~\& Quinn, H.~R.~CP Conservation in the Presence of Pseudoparticles. \emph{Phys.~Rev.~Lett.~}\textbf{38,} 1440 (1977).

\bibitem{Peccei1977b} Peccei, R.~D.~\& Quinn, H.~R.~Constraints imposed by CP conservation in the presence of pseudoparticles. \emph{Phys.~Rev.~D~}\textbf{16,} 1791 (1977).

\bibitem{Kim1979} J.~E.~Kim, Weak-Interaction Singlet and Strong CP Invariance. \emph{Phys.~Rev.~Lett.~}\textbf{43}, 103 (1979).

\bibitem{Shifman1980} M. A. Shifman, A. I. Vainshtein, and V. I. Zakharov, Can confinement ensure natural CP invariance of strong interactions? \emph{Nucl.~Phys.~B} \textbf{166}, 493 (1980).

\bibitem{Zhitnitsky1980} A.~R.~Zhitnitsky, The Weinberg Model Of The CP Violation And T Odd Correlations In Weak Decays. \emph{Yad.~Fiz.}~\textbf{31}, 1024 (1980) [\emph{Sov.~J.~Nucl.~Phys.}~\textbf{31}, 529 (1980)].

\bibitem{Dine1981} M. Dine, W. Fischler, and M. Srednicki, A simple solution to the strong CP problem with a harmless axion. \emph{Phys.~Lett.~B} \textbf{104}, 199 (1981).

\bibitem{Sikivie2009} Sikivie, P.~\& Yang, Q.~Bose-Einstein Condensation of Dark Matter Axions. \emph{Phys.~Rev.~Lett.~}\textbf{103,} 111301 (2009).



\bibitem{Sikivie1983} P.~Sikivie, Experimental Tests of the "Invisible" Axion. \emph{Phys.~Rev.~Lett.}~51, 1415 (1983).

\bibitem{ADMX2010} Asztalos, S.~J.~\emph{et.~al.~}SQUID-Based Microwave Cavity Search for Dark-Matter Axions. \emph{Phys.~Rev.~Lett.~}\textbf{104,} 041301 (2010).

\bibitem{Graham2011} P.~W.~Graham and S.~Rajendran, Axion dark matter detection with cold molecules. \emph{Phys.~Rev.~D} \textbf{84}, 055013 (2011).

\bibitem{Graham2013} P.~W.~Graham and S.~Rajendran, New observables for direct detection of axion dark matter. \emph{Phys.~Rev.~D} \textbf{88}, 035023 (2013).

\bibitem{Stadnik2014axions} Stadnik, Y.~V.~\& Flambaum, V.~V.~Axion-induced effects in atoms, molecules, and nuclei: Parity nonconservation, anapole moments, electric dipole moments, and spin-gravity and spin-axion momentum couplings. \emph{Phys.~Rev.~D} \textbf{89,} 043522 (2014).

\bibitem{CASPER2014} Budker, D., Graham, P.~W., Ledbetter, M., Rajendran, S.~\& Sushkov, A.~O.~Proposal for a Cosmic Axion Spin Precession Experiment (CASPEr). \emph{Phys.~Rev.~X} \textbf{4,} 021030 (2014).

\bibitem{Roberts2014prl} B.~M.~Roberts, Y.~V.~Stadnik, V.~A.~Dzuba, V.~V.~Flambaum, N.~Leefer, and D.~Budker. Limiting P-Odd Interactions of Cosmic Fields with Electrons, Protons, and Neutrons. \emph{Phys.~Rev.~Lett.}~\textbf{113}, 081601 (2014).

\bibitem{Roberts2014long} B.~M.~Roberts, Y.~V.~Stadnik, V.~A.~Dzuba, V.~V.~Flambaum, N.~Leefer, and D.~Budker. Parity-violating interactions of cosmic fields with atoms, molecules, and nuclei: Concepts and calculations for laboratory searches and extracting limits. \emph{Phys.~Rev.~D} \textbf{90,} 096005 (2014).

\bibitem{Derevianko2014} Derevianko, A.~\& Pospelov, M.~Hunting for topological dark matter with atomic clocks. \emph{Nature Physics} \textbf{10,} 933-936 (2014).

\bibitem{Olive2008} Olive, K.~A.~\& Pospelov, M.~Environmental dependence of masses and coupling constants. \emph{Phys.~Rev.~D} \textbf{77,} 043524 (2008).



\bibitem{Vilenkin1985} Vilenkin, A.~Cosmic strings and domain walls. \emph{Phys.~Rep.~}\textbf{121,} 263-315 (1985).

\bibitem{Marsh2015soliton} D.~J.~E.~Marsh and A.-R.~Pop,~Axion dark matter, solitons, and the cusp-core problem. arXiv:1502.03456.


\bibitem{GNOME2013} Pospelov, M.~\emph{et.~al.}~Detecting domain walls of axionlike models using terrestrial experiments. \emph{Phys.~Rev.~Lett.}~\textbf{110,} 021803 (2013).

\bibitem{Stadnik2014defects} Stadnik, Y.~V.~\& Flambaum, V.~V.~Searching for Topological Defect Dark Matter via Nongravitational Signatures. \emph{Phys.~Rev.~Lett.}~\textbf{113,} 151301 (2014).



\bibitem{Hinkley2013} Hinkley, N.~\emph{et.~al.}~An Atomic Clock with $10^{–18}$ Instability. \emph{Science} \textbf{341,} 1215-1218 (2013).

\bibitem{Bloom2014} Bloom, B.~J.~\emph{et.~al.}~An optical lattice clock with accuracy and stability at the $10^{–18}$ level. \emph{Nature (London)} \textbf{506,} 71-75 (2014).

\bibitem{Jefferts2013} Jefferts, S.~R.~\emph{Atomic Clocks: Primary Frequency Standards at NIST} (8th Annual DOE Laser Safety Officer Workshop, Menlo Park, 2013).




\bibitem{Kostelecky2011RMP} Kosteleck\'y, V.~A.~\& Russell, N.~Data tables for Lorentz and CPT violation. \emph{Rev.~Mod.~Phys.~}\textbf{83,} 11 (2011).





\bibitem{LIGOBook2014} Massimo, B.~(Ed.) \emph{Advanced Interferometers and the Search for Gravitational Waves: Lectures from the First VESF School on Advanced Detectors for Gravitational Waves}. (Springer, Cham, 2014).

\bibitem{VanTilburg2015} A.~Arvanitaki, J.~Huang, and K.~Van Tilburg, Searching for dilaton dark matter with atomic clocks. \emph{Phys.~Rev.~D} \textbf{91}, 015015 (2015).



\bibitem{Holland2009} D.~Meiser, J.~Ye, D.~R.~Carlson, and M.~J.~Holland, Prospects for a Millihertz-Linewidth Laser. \emph{Phys.~Rev.~Lett.}~\textbf{102}, 163601 (2009).

\bibitem{Thomson2012a} J.~G.~Bohnet, Z.~Chen, J.~M.~Weiner, D.~Meiser, M.~J.~Holland and J.~K.~Thompson. A steady-state superradiant laser with less than one intracavity photon. \emph{Nature (London)} \textbf{484}, 78 (2012).

\bibitem{Thomson2012b} J.~G.~Bohnet, Z.~Chen, J.~M.~Weiner, K.~C.~Cox, and J.~K.~Thompson. Relaxation Oscillations, Stability, and Cavity Feedback in a Superradiant Raman Laser. \emph{Phys.~Rev.~Lett.}~\textbf{109}, 253602 (2012).


\bibitem{Tedesco2006} V.~V.~Flambaum and A.~F.~Tedesco, Dependence of nuclear magnetic moments on quark masses and limits on temporal variation of fundamental constants from atomic clock experiments. \emph{Phys.~Rev.~C} \textbf{73}, 055501 (2006).

\bibitem{Flambaum1999Theor} V.~A.~Dzuba, V.~V.~Flambaum, and J.~K.~Webb, Calculations of the relativistic effects in many-electron atoms and space-time variation of fundamental constants. \emph{Phys.~Rev.~A} \textbf{59}, 230 (1999).

\bibitem{Dinh2009} T.~H.~Dinh, A.~Dunning, V.~A.~Dzuba, and V.~V.~Flambaum, The sensitivity of hyperfine structure to nuclear radius and quark mass variation. \emph{Phys.~Rev.~A} \textbf{79}, 054102 (2009).




\bibitem{Gill2014} Godun, R.~M.~\emph{et.~al.~}Frequency Ratio of Two Optical Clock Transitions in $^{171}$Yb$^+$ and Constraints on the Time Variation of Fundamental Constants. \emph{Phys.~Rev.~Lett.~}\textbf{113,} 210801 (2014).

\bibitem{Peik2014} Huntemann, N.~\emph{et.~al.~}Improved Limit on a Temporal Variation of $m_p/m_e$ from Comparisons of Yb$^+$ and Cs Atomic Clocks. \emph{Phys.~Rev.~Lett.~}\textbf{113,} 210802 (2014).






\bibitem{Ubachs2006} Reinhold, E.~\emph{et.~al.~}Indication of a Cosmological Variation of the Proton-Electron Mass Ratio Based on Laboratory Measurement and Reanalysis of H$_2$ Spectra. \emph{Phys.~Rev.~Lett.~}\textbf{96,} 151101 (2006).

\bibitem{Flambaum1999} Dzuba, V.~A., Flambaum, V.~V.~\& Webb, J.~K.~Space-Time Variation of Physical Constants and Relativistic Corrections in Atoms. \emph{Phys.~Rev.~Lett.~}\textbf{82,} 888 (1999).



\bibitem{NIST-database} NIST, Computational Chemistry Comparison and Benchmark DataBase, \url{http://cccbdb.nist.gov/default.htm}


\bibitem{TAMA2004} TAMA Collaboration, Observation results by the TAMA300 detector on gravitational wave bursts from stellar-core collapses. Phys.~Rev.~D \textbf{71,} 082002 (2005).

\bibitem{LIGO2012} Ligo Scientific Collaboration and Virgo Collaboration, Sensitivity Achieved by the LIGO and Virgo Gravitational Wave Detectors during
LIGO's Sixth and Virgo's Second and Third Science Runs. arXiv:1203.2674.

\bibitem{GEO2014} K.~L.~Dooley, Status of GEO600. arXiv:1411.6588.

\bibitem{FHolometer} A.~Chou \emph{et.~al.~}The Fermilab Holometer: A program to measure Planck scale indeterminacy, Fermilab publication, 2009.






















\end{thebibliography}


\appendix

\section*{Supplemental material}

\begin{figure}[h!]
\begin{center}
\includegraphics[width=8cm]{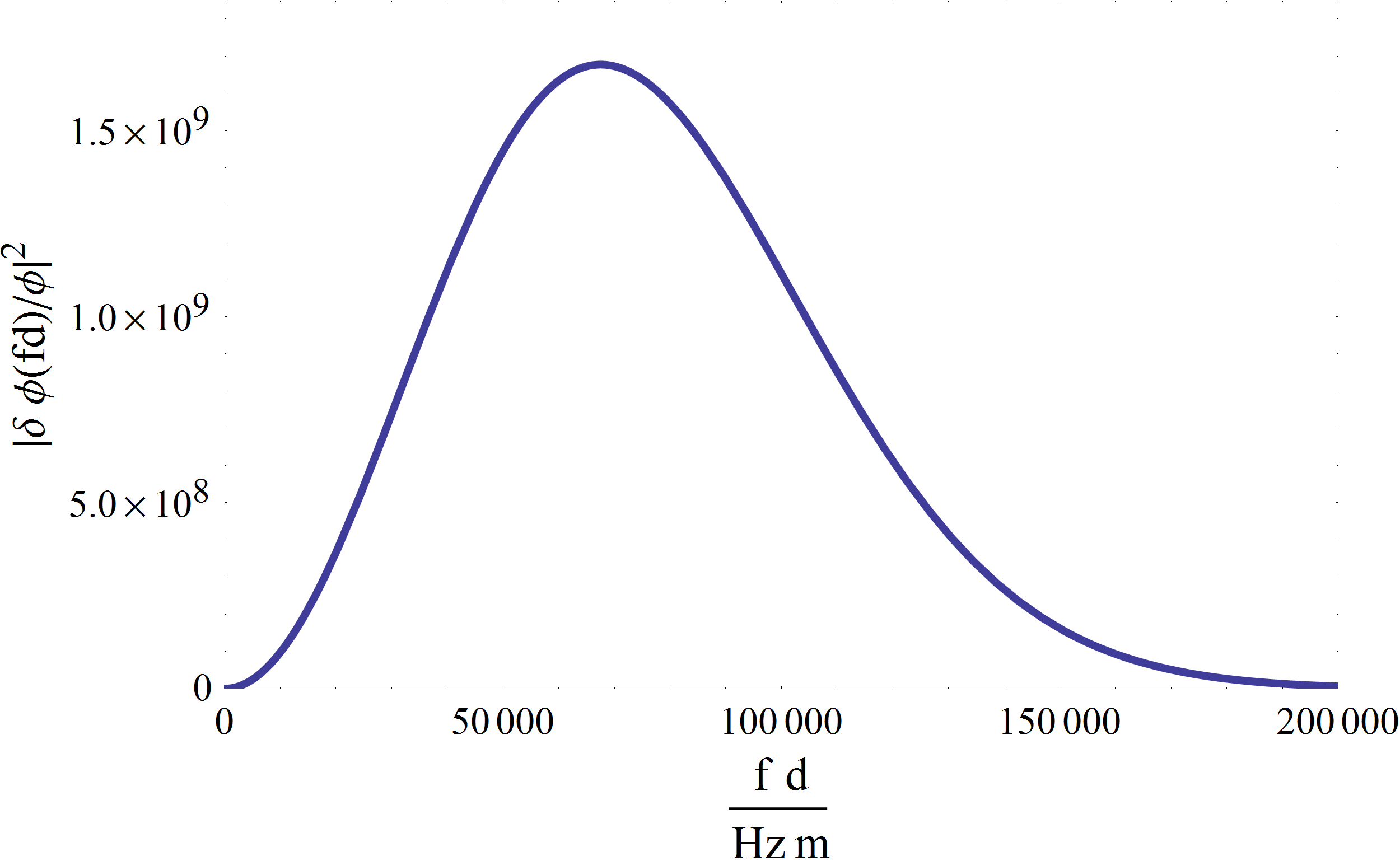}
\caption{(Color online) Power spectrum [in units $ \rho_{\textrm{TDM}}^2 L^2 \tau^2 d^2 \hbar^2 c^2 \pi^3 / (\Lambda_X')^4 v^2 \times $(m/s)$^2$] versus $fd$, produced by a domain wall with a Gaussian cross-sectional profile passing directly along one of the arms of a Michelson-Morley interferometer with arms of equal length $L$.} 
\label{fig:d_larger_cf_L}
\end{center}
\end{figure}

\begin{figure*}[h!]
\begin{center}
\includegraphics[width=5.7cm]{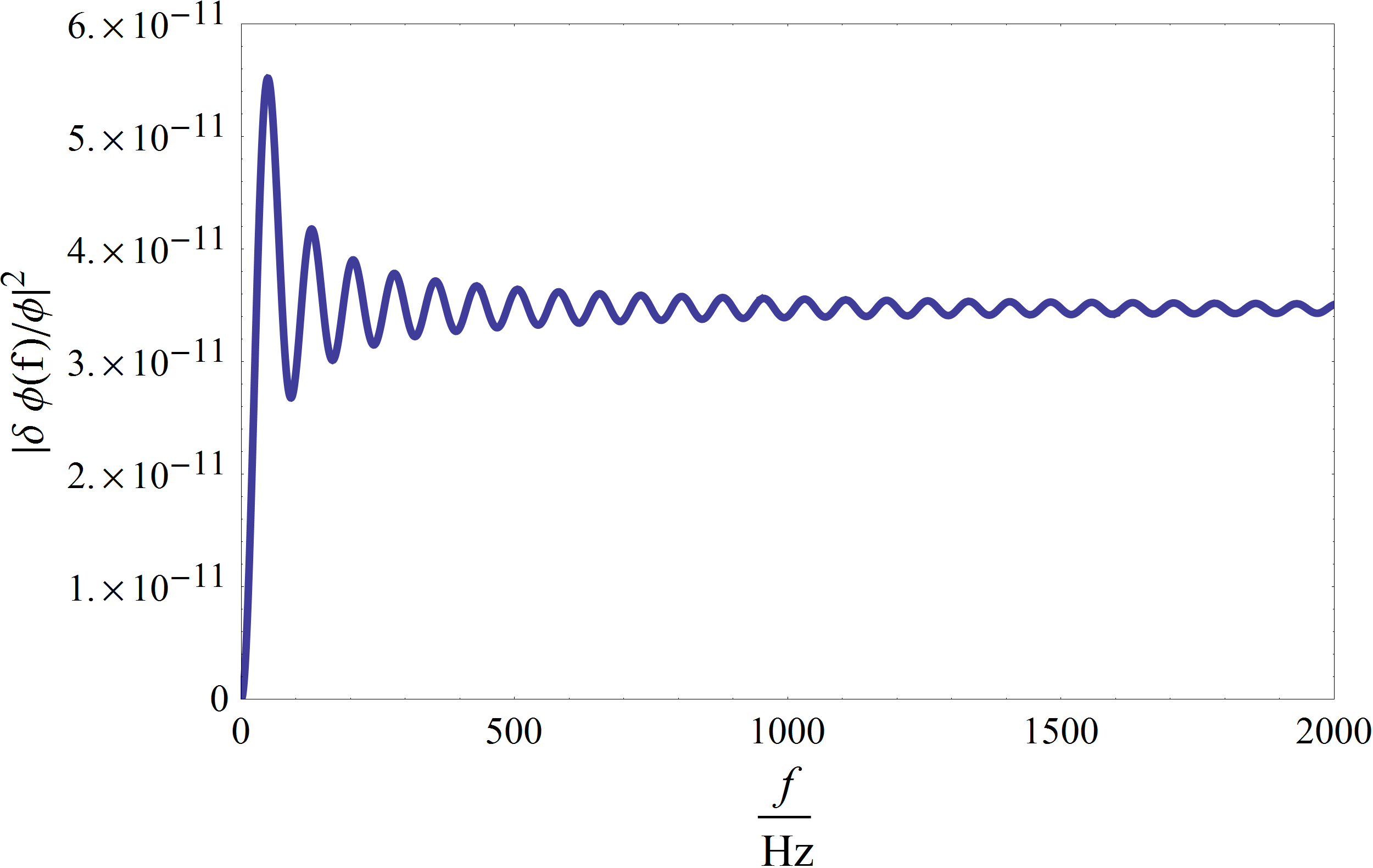}
\includegraphics[width=5.7cm]{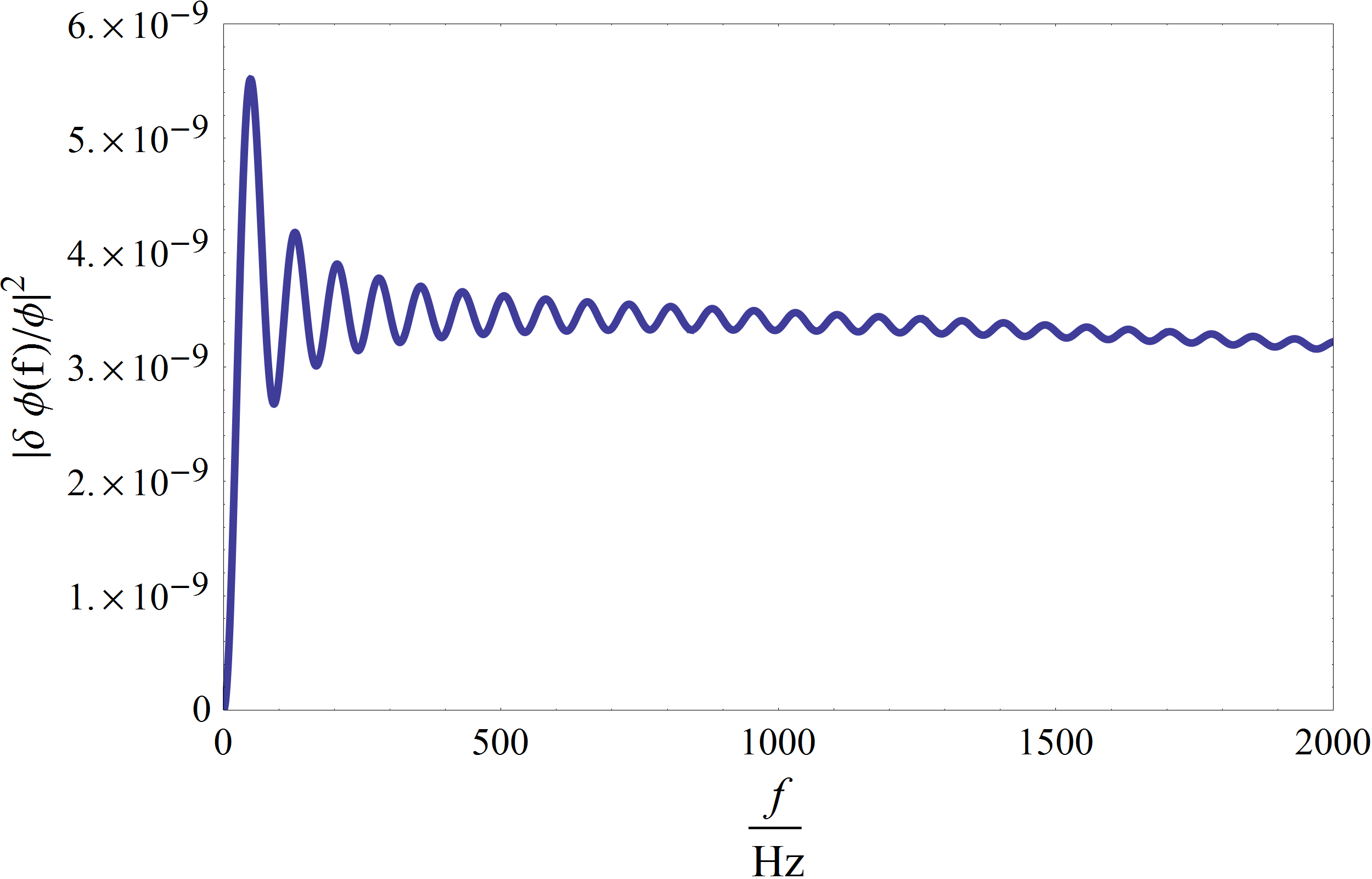}
\includegraphics[width=5.7cm]{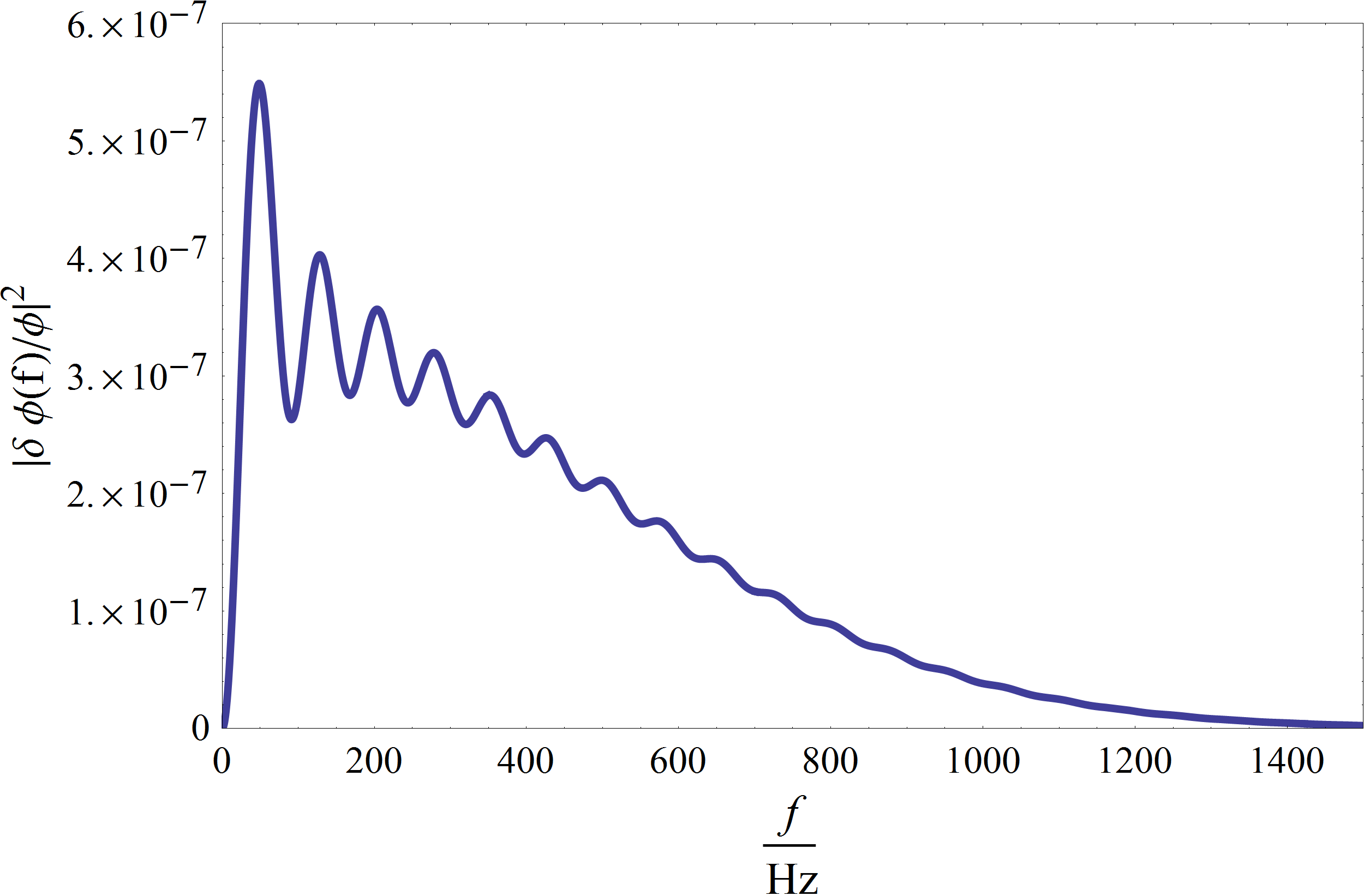}
\includegraphics[width=5.7cm]{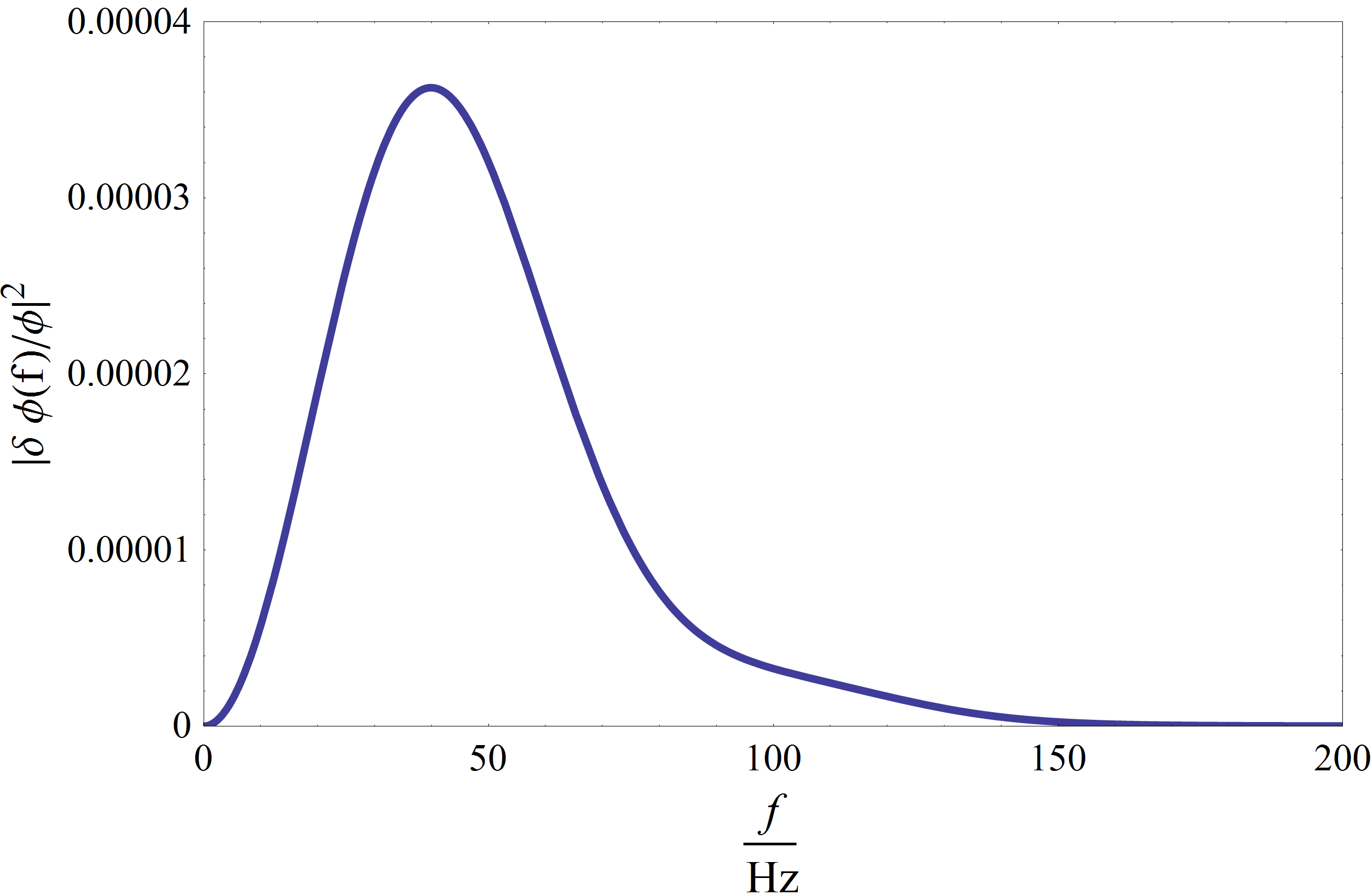}
\includegraphics[width=5.7cm]{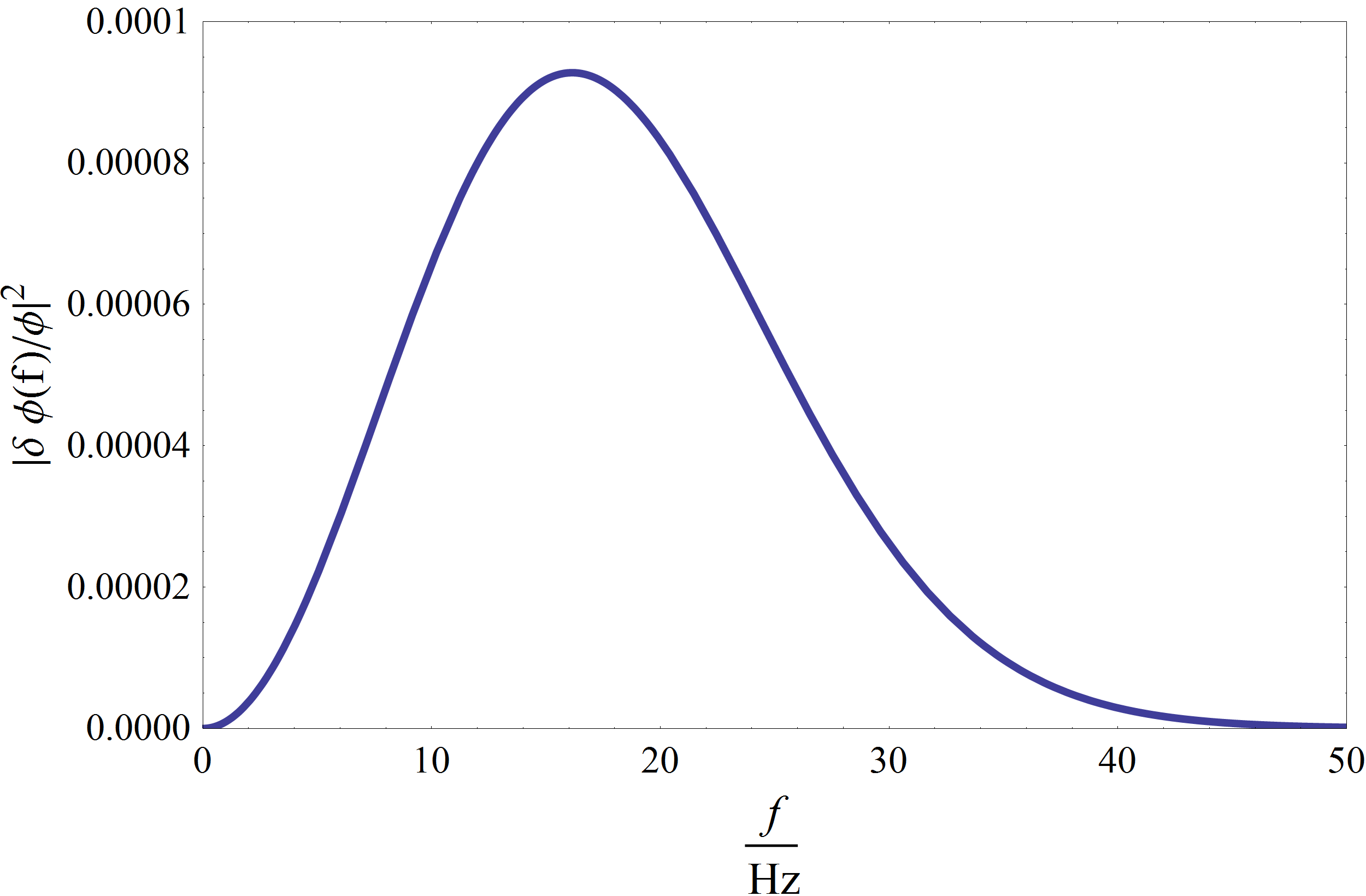}
\caption{(Color online) Power spectra [in units $ \rho_{\textrm{TDM}}^2 \tau^2 v^2 d^2 \hbar^2 c^2 / (\Lambda_X')^4 \times $s$^2$] versus frequency, produced by a domain wall with a Gaussian cross-sectional profile passing directly along one of the arms of a LIGO interferometer ($L=4$ km). From left to right: $d=1$ m, $d=10$ m, $d=100$ m, $d=1000$ m, $d=4000$ m.} 
\label{fig:LIGO}
\end{center}
\end{figure*}

\begin{figure*}[h!]
\begin{center}
\includegraphics[width=5.7cm]{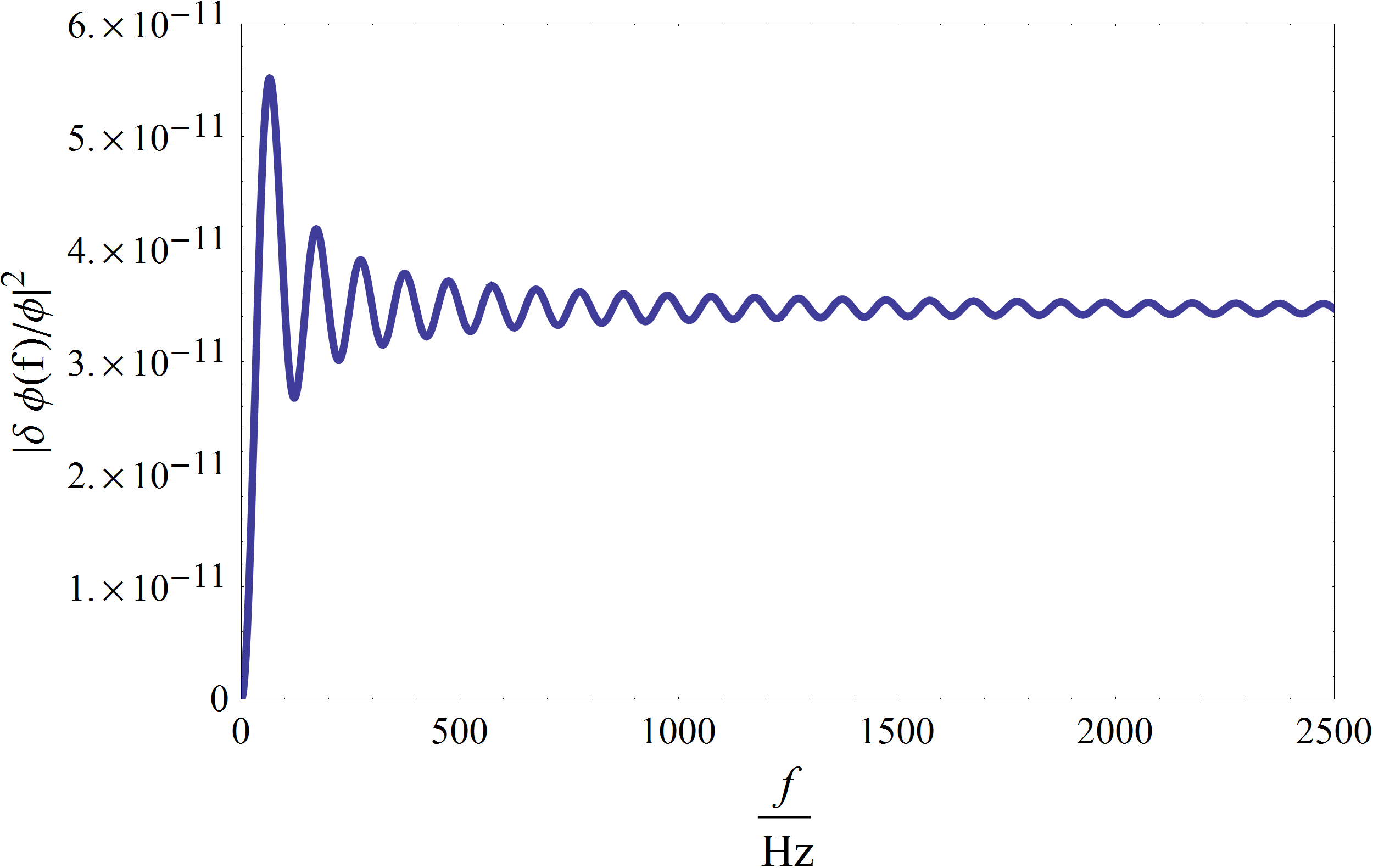}
\includegraphics[width=5.7cm]{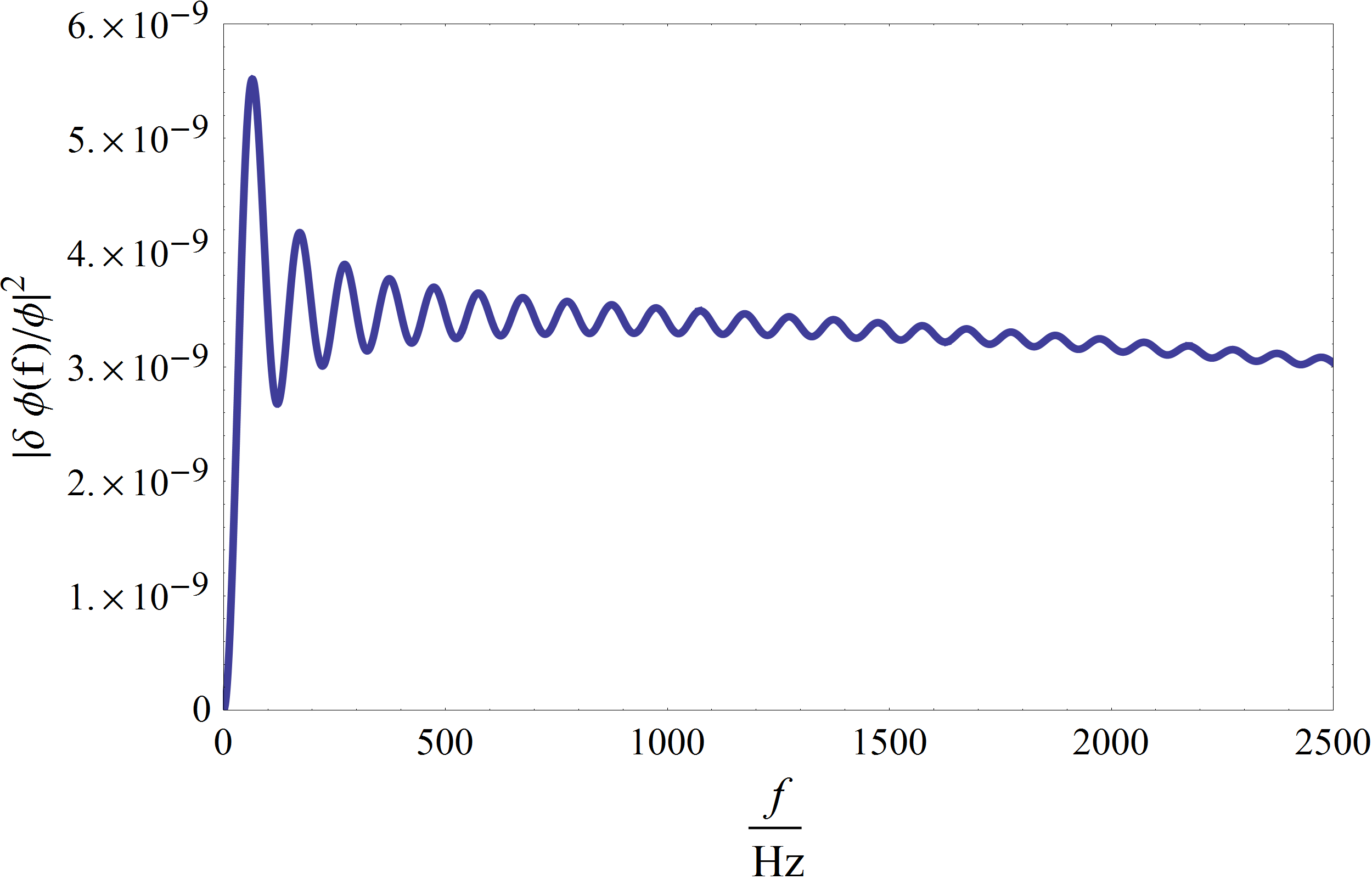}
\includegraphics[width=5.7cm]{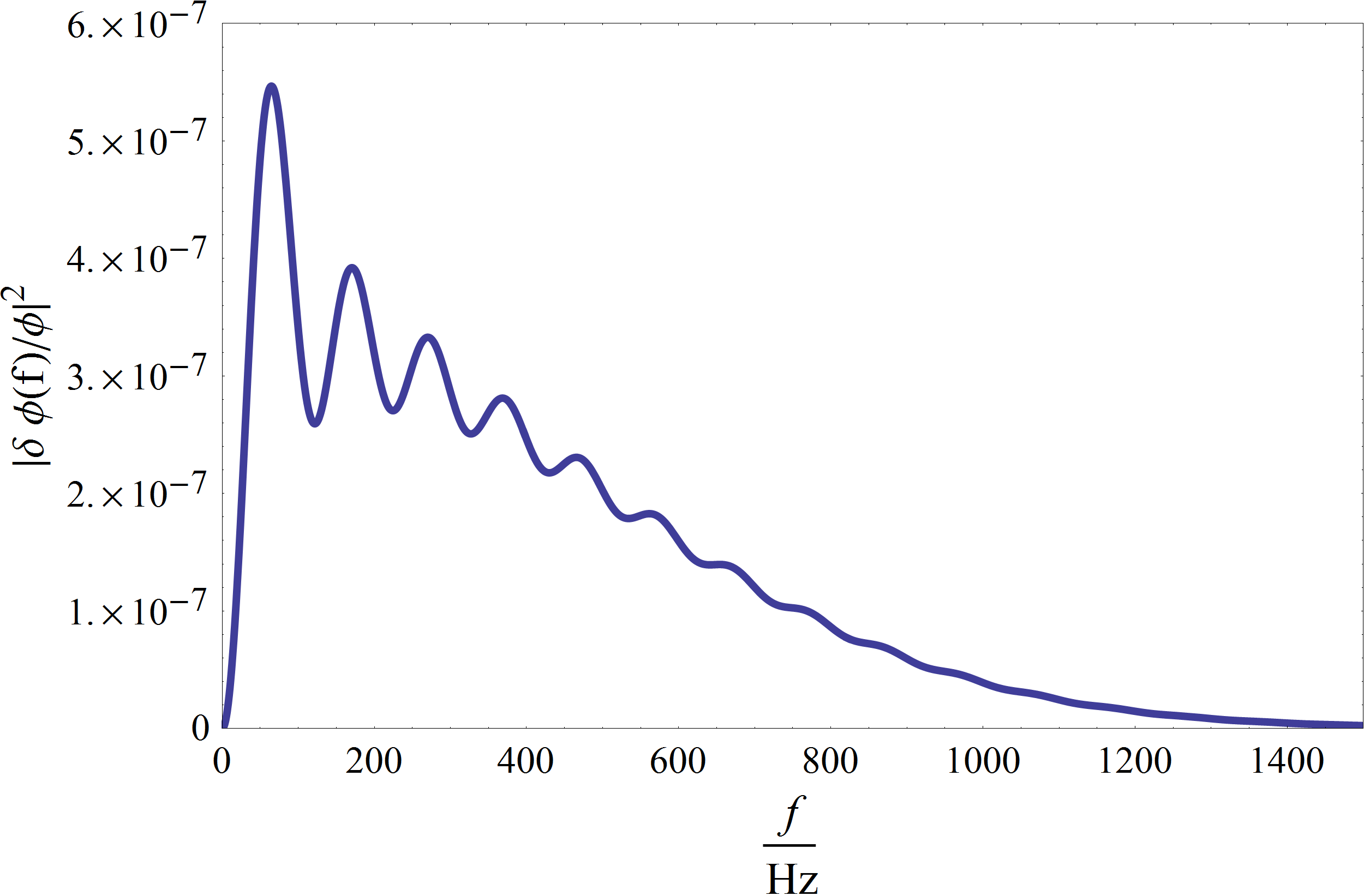}
\includegraphics[width=5.7cm]{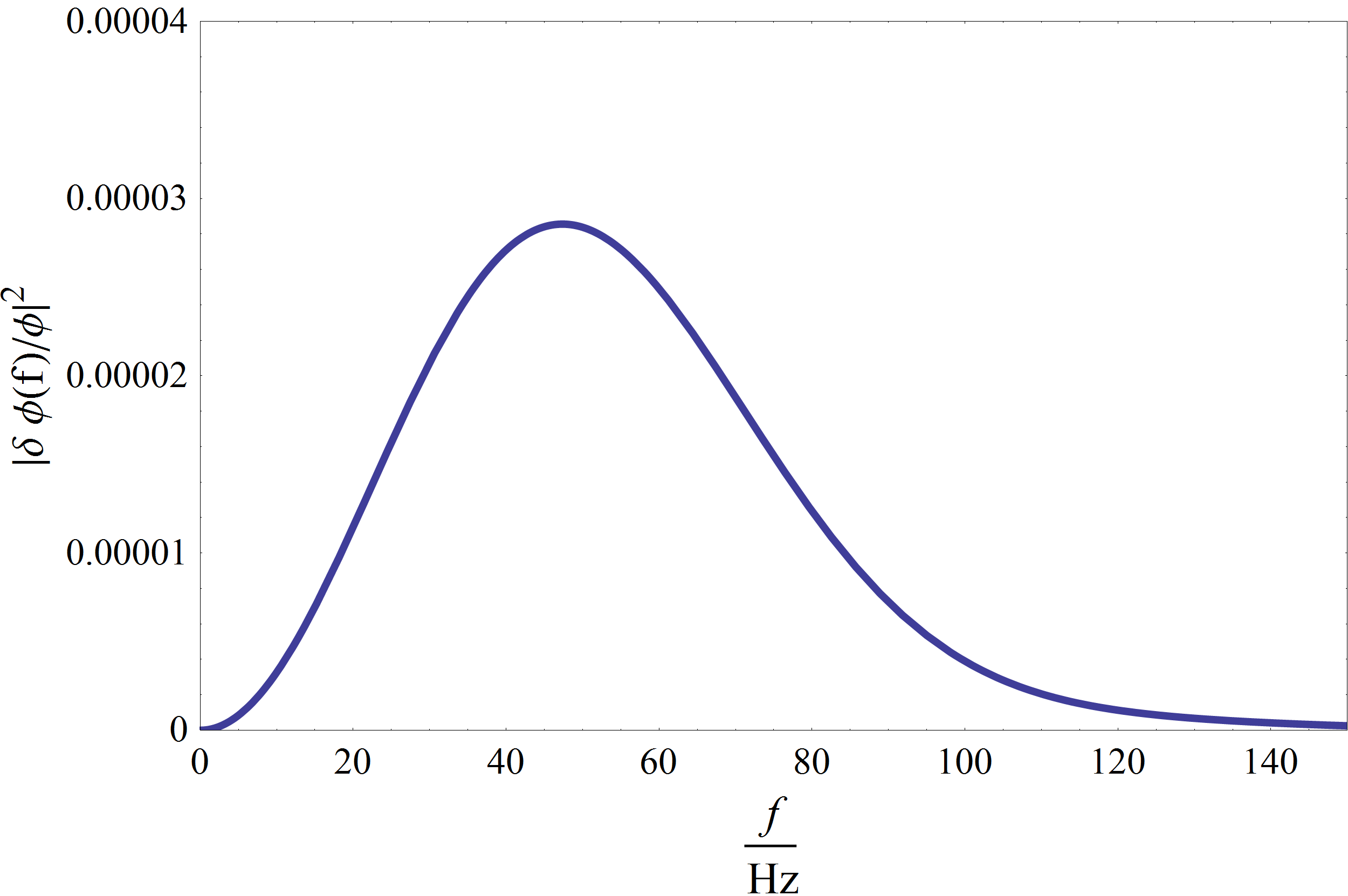}
\includegraphics[width=5.7cm]{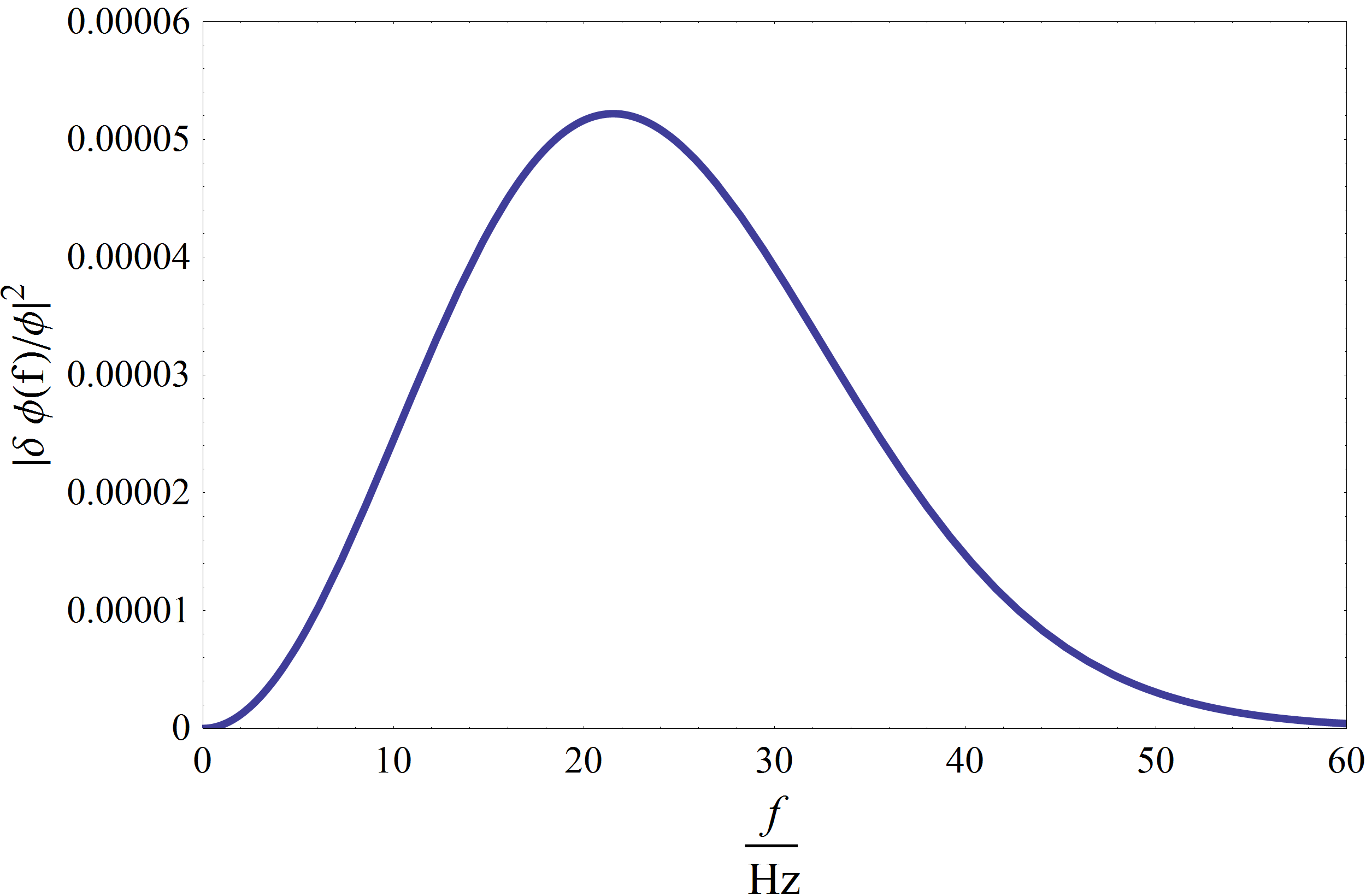}
\caption{(Color online) Power spectra [in units $ \rho_{\textrm{TDM}}^2 \tau^2 v^2 d^2 \hbar^2 c^2 / (\Lambda_X')^4 \times $s$^2$] versus frequency, produced by a domain wall with a Gaussian cross-sectional profile passing directly along one of the arms of a Virgo interferometer ($L=3$ km). From left to right: $d=1$ m, $d=10$ m, $d=100$ m, $d=1000$ m, $d=3000$ m.} 
\label{fig:Virgo}
\end{center}
\end{figure*}

\begin{figure*}[h!]
\begin{center}
\includegraphics[width=7cm]{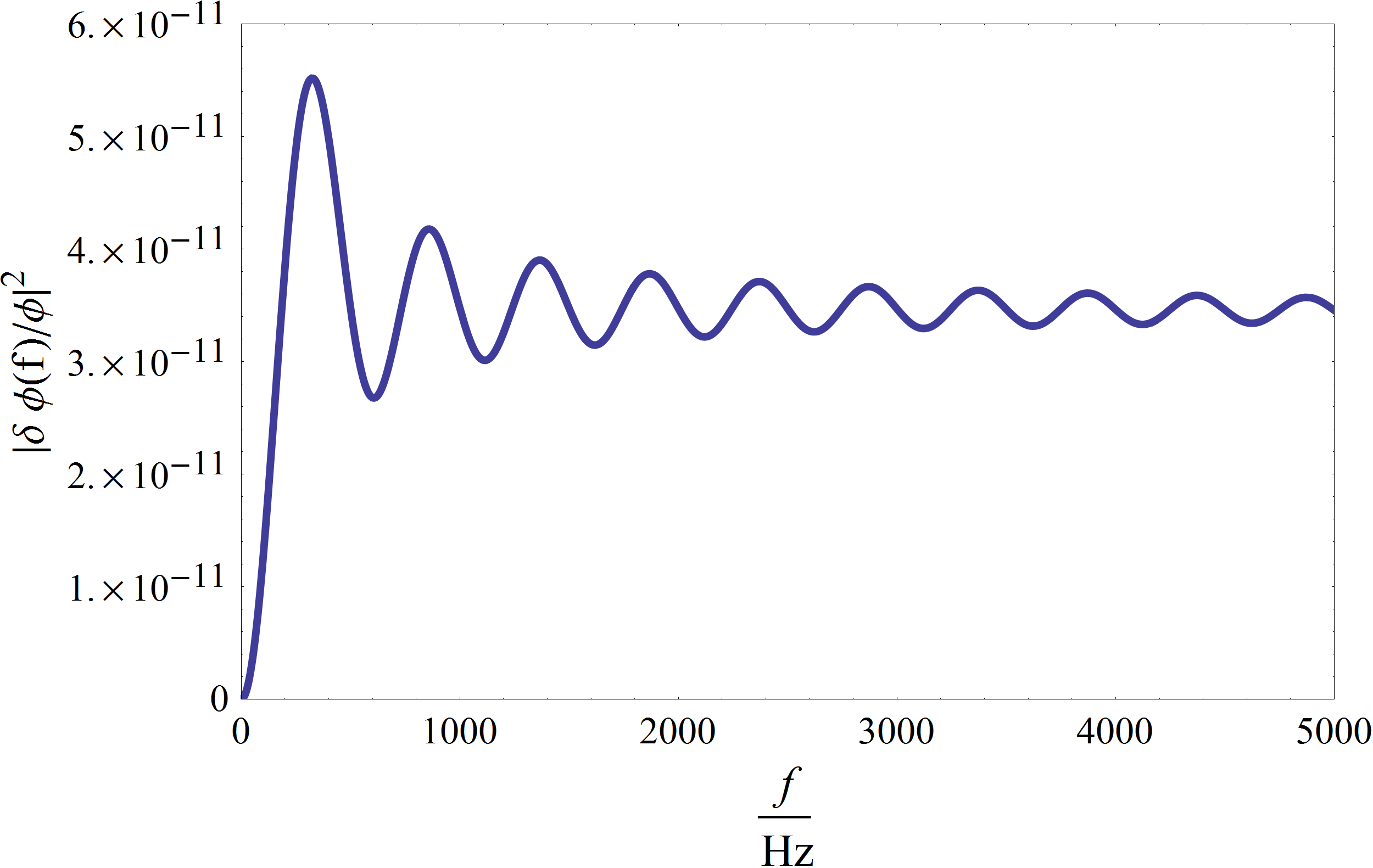}
\includegraphics[width=7cm]{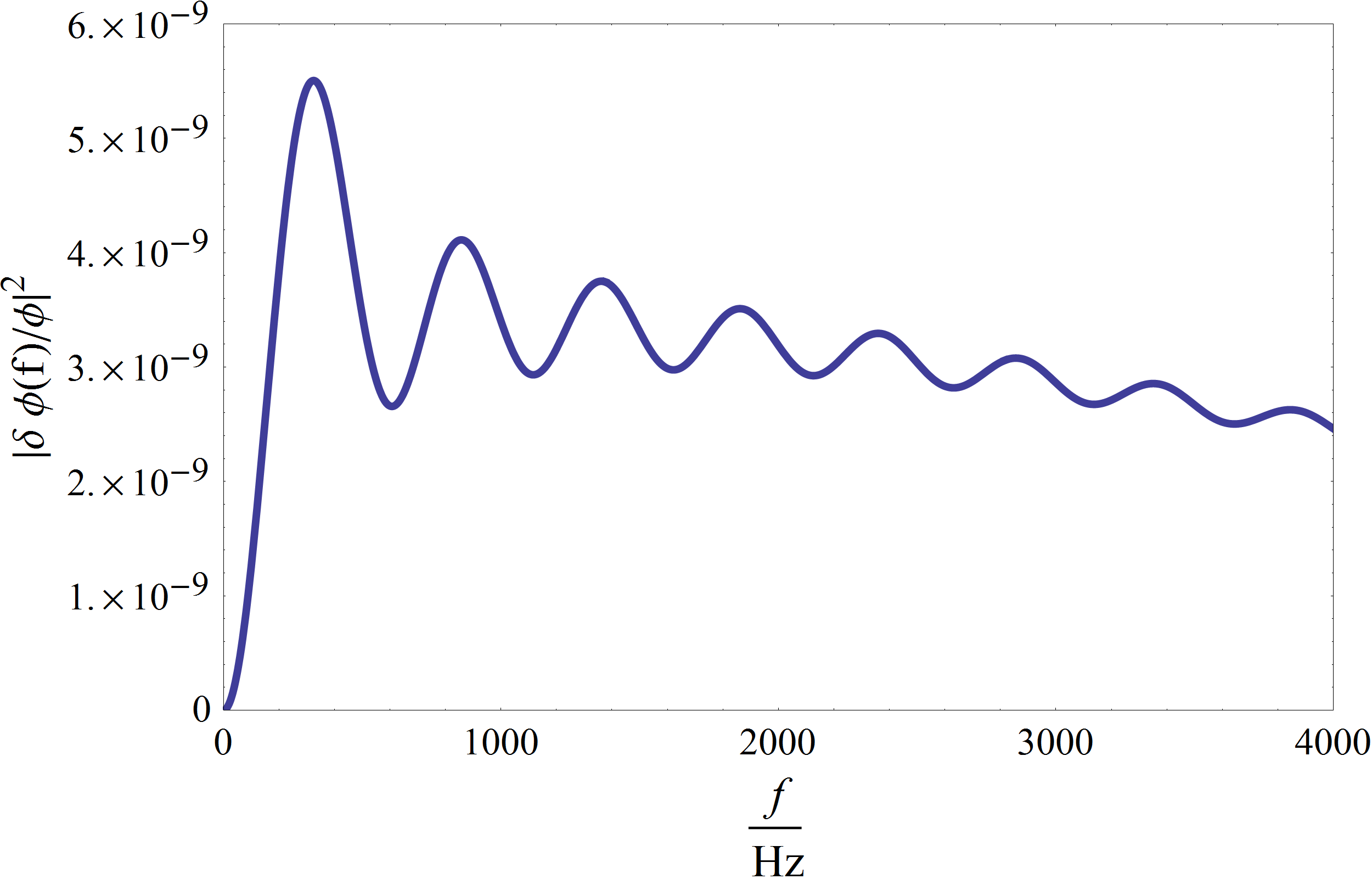}
\includegraphics[width=7cm]{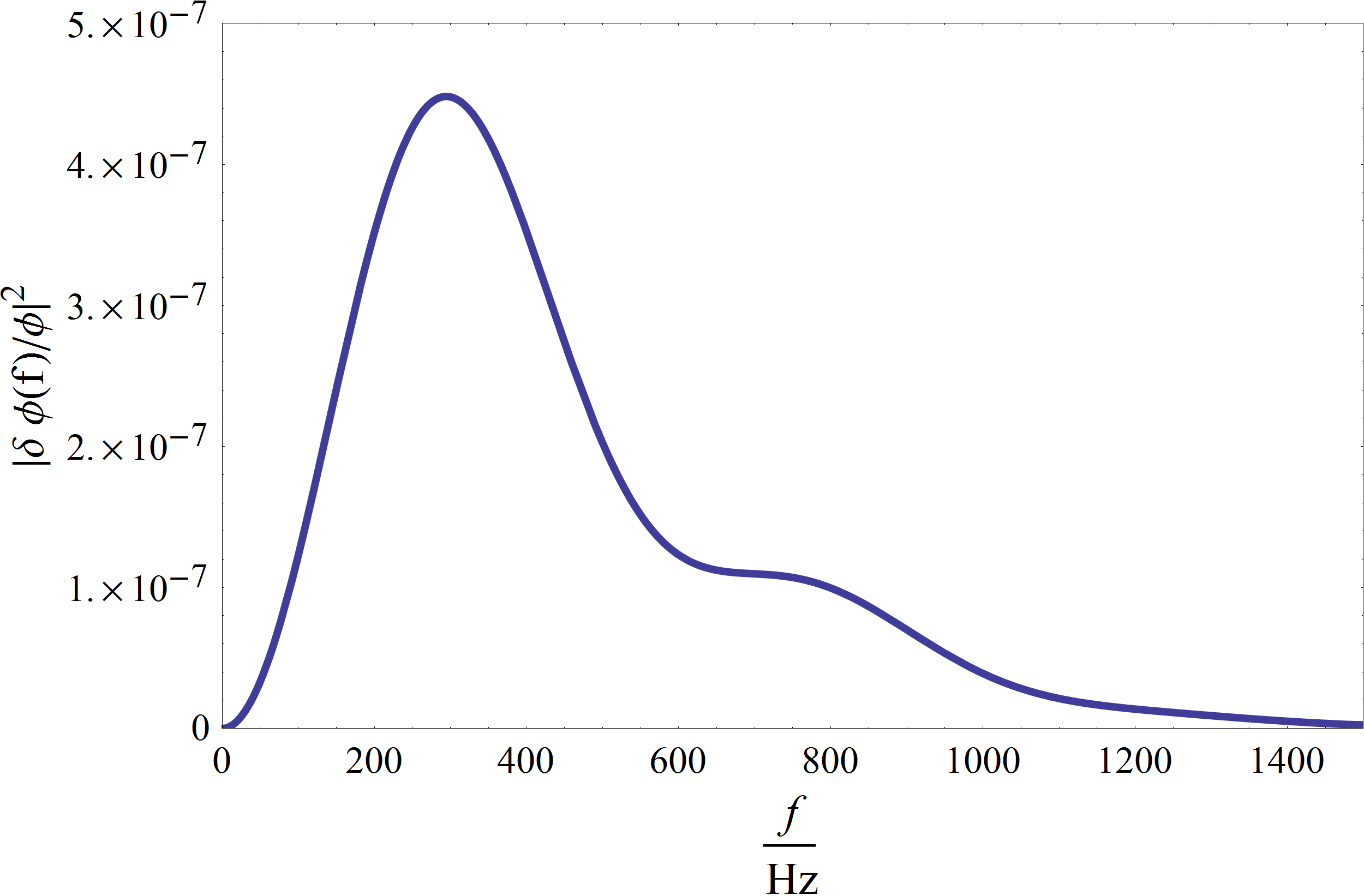}
\includegraphics[width=7cm]{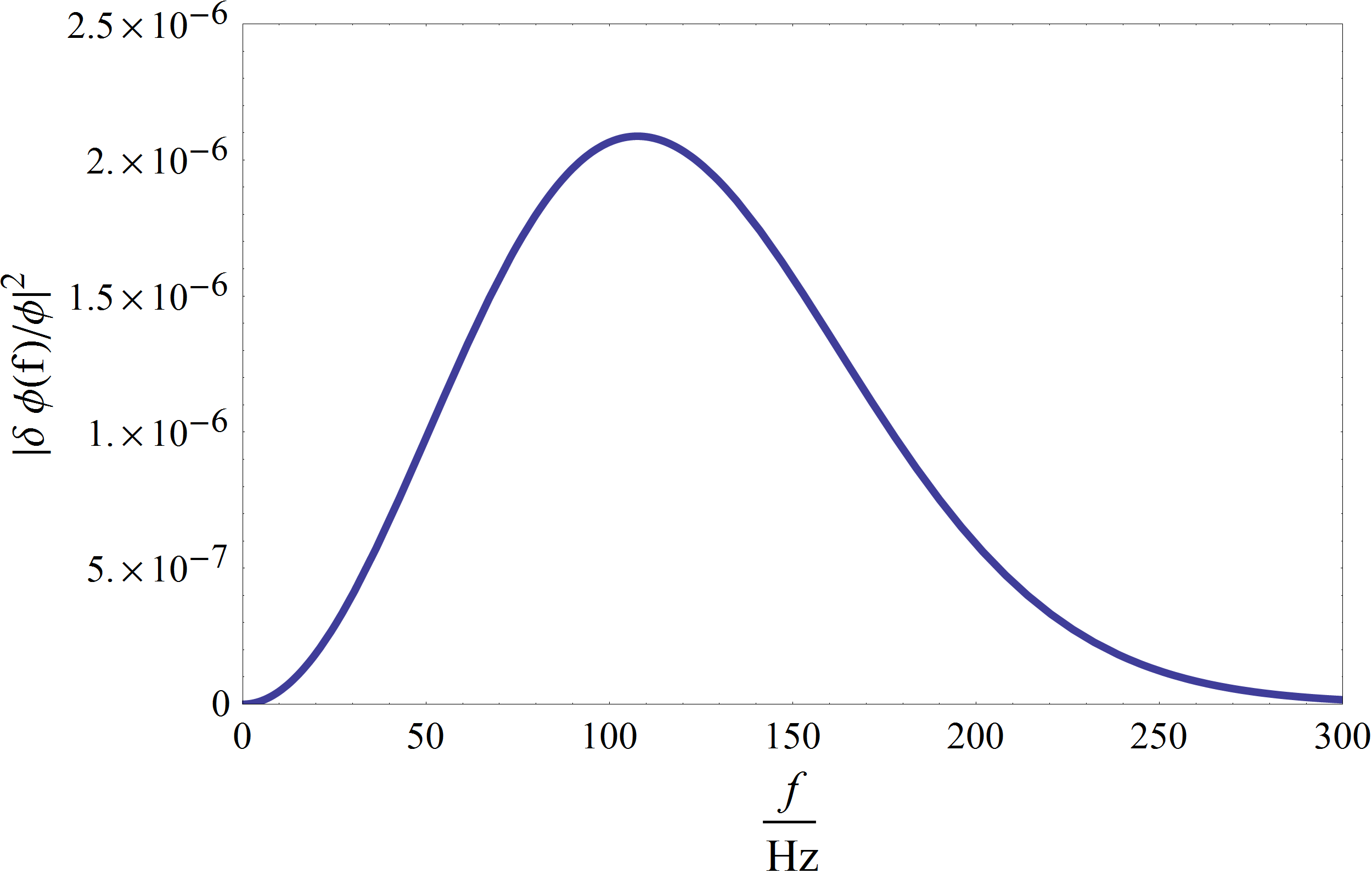}
\caption{(Color online) Power spectra [in units $ \rho_{\textrm{TDM}}^2 \tau^2 v^2 d^2 \hbar^2 c^2 / (\Lambda_X')^4 \times $s$^2$] versus frequency, produced by a domain wall with a Gaussian cross-sectional profile passing directly along one of the arms of a GEO600 interferometer ($L=600$ m). From left to right: $d=1$ m, $d=10$ m, $d=100$ m, $d=600$ m.} 
\label{fig:GEO}
\end{center}
\end{figure*}

\begin{figure*}[h!]
\begin{center}
\includegraphics[width=7cm]{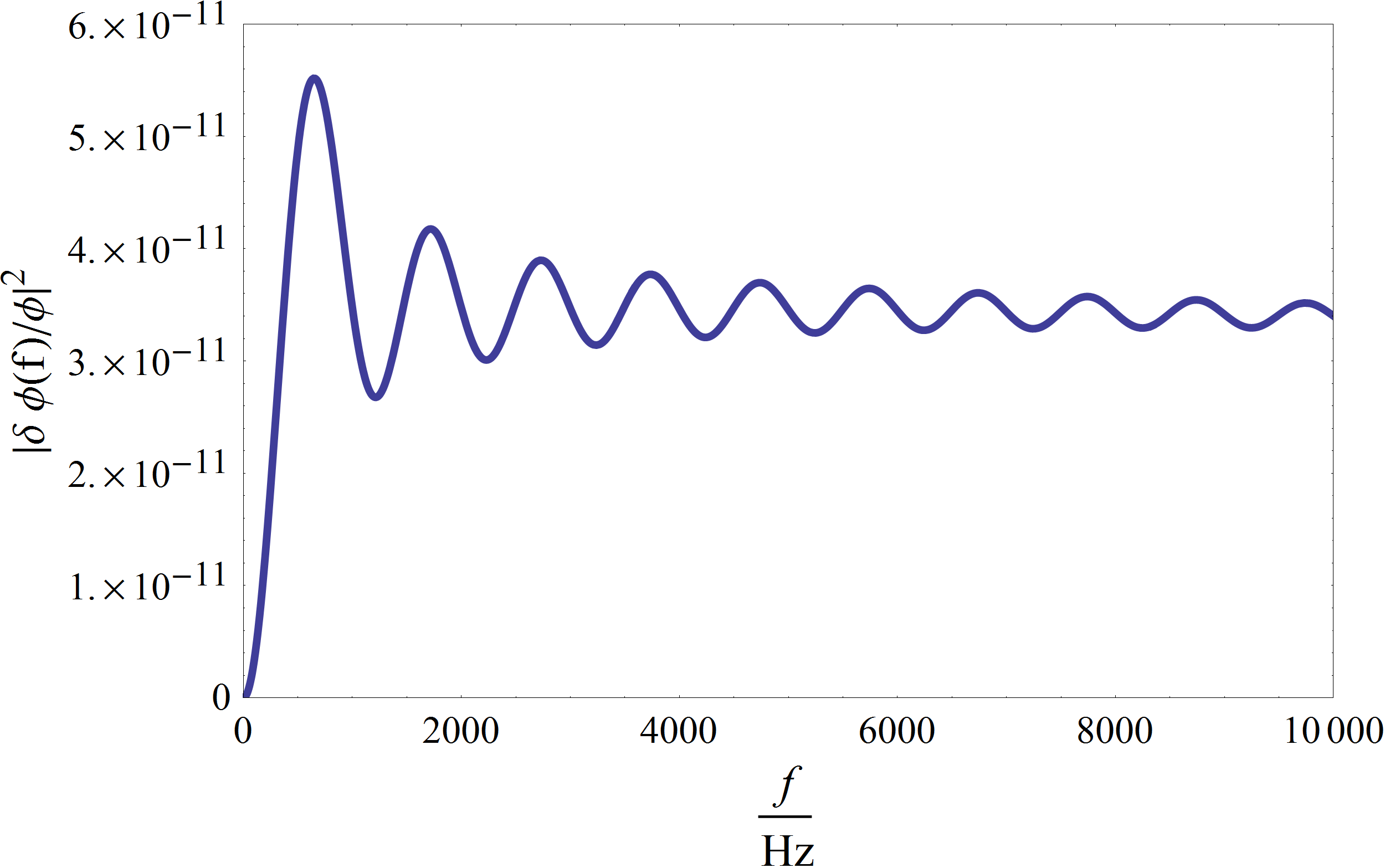}
\includegraphics[width=7cm]{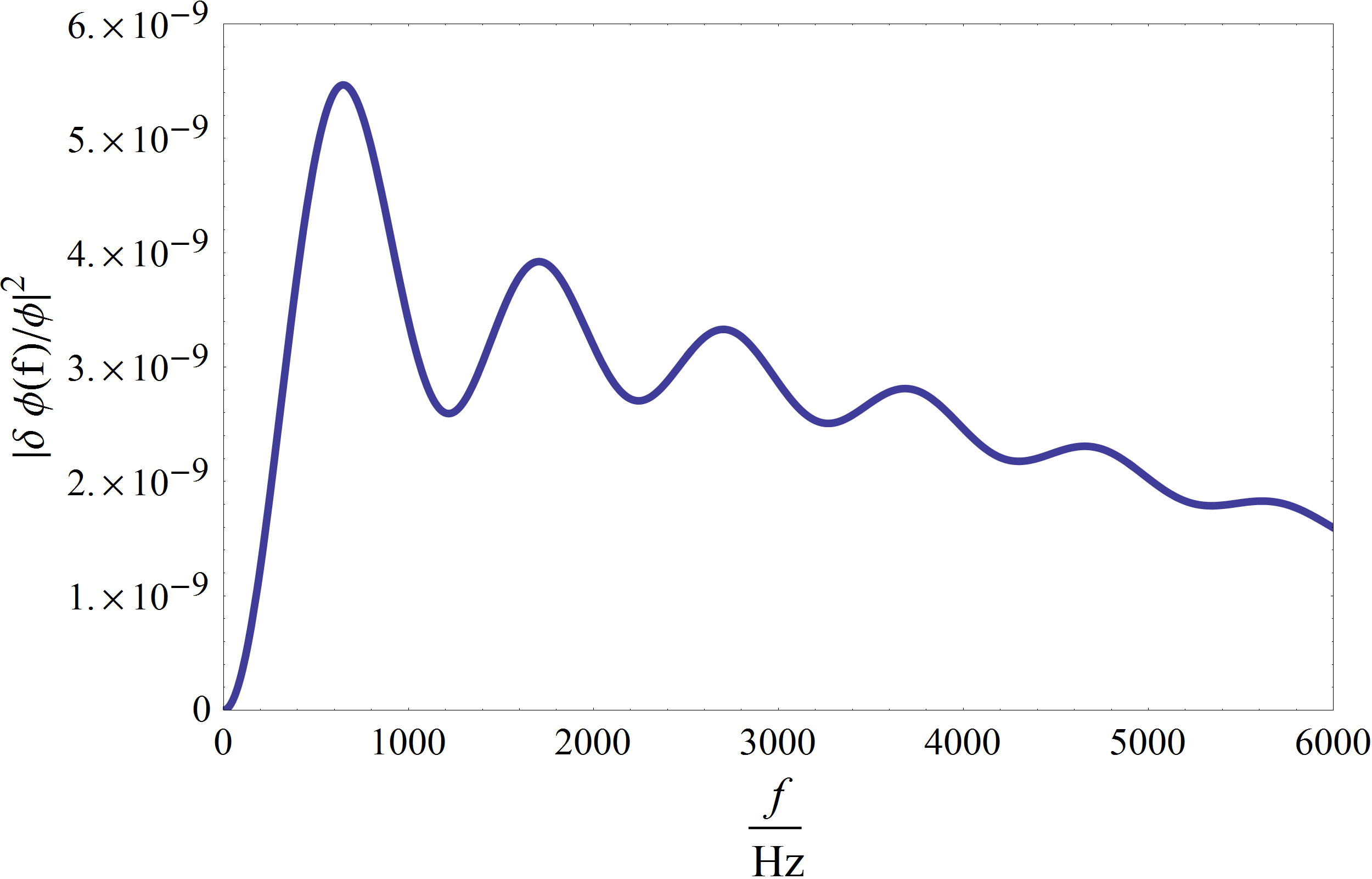}
\includegraphics[width=7cm]{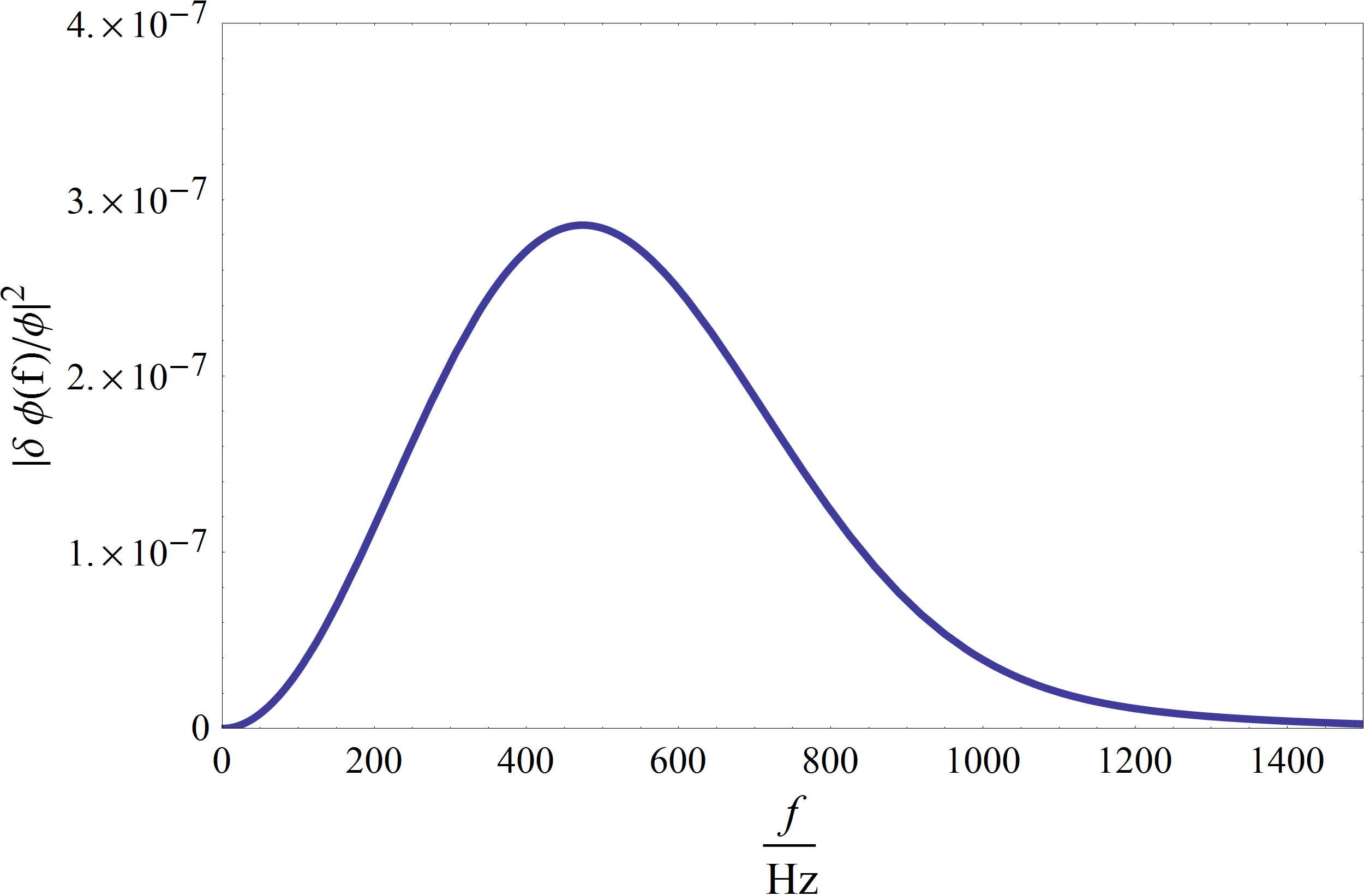}
\includegraphics[width=7cm]{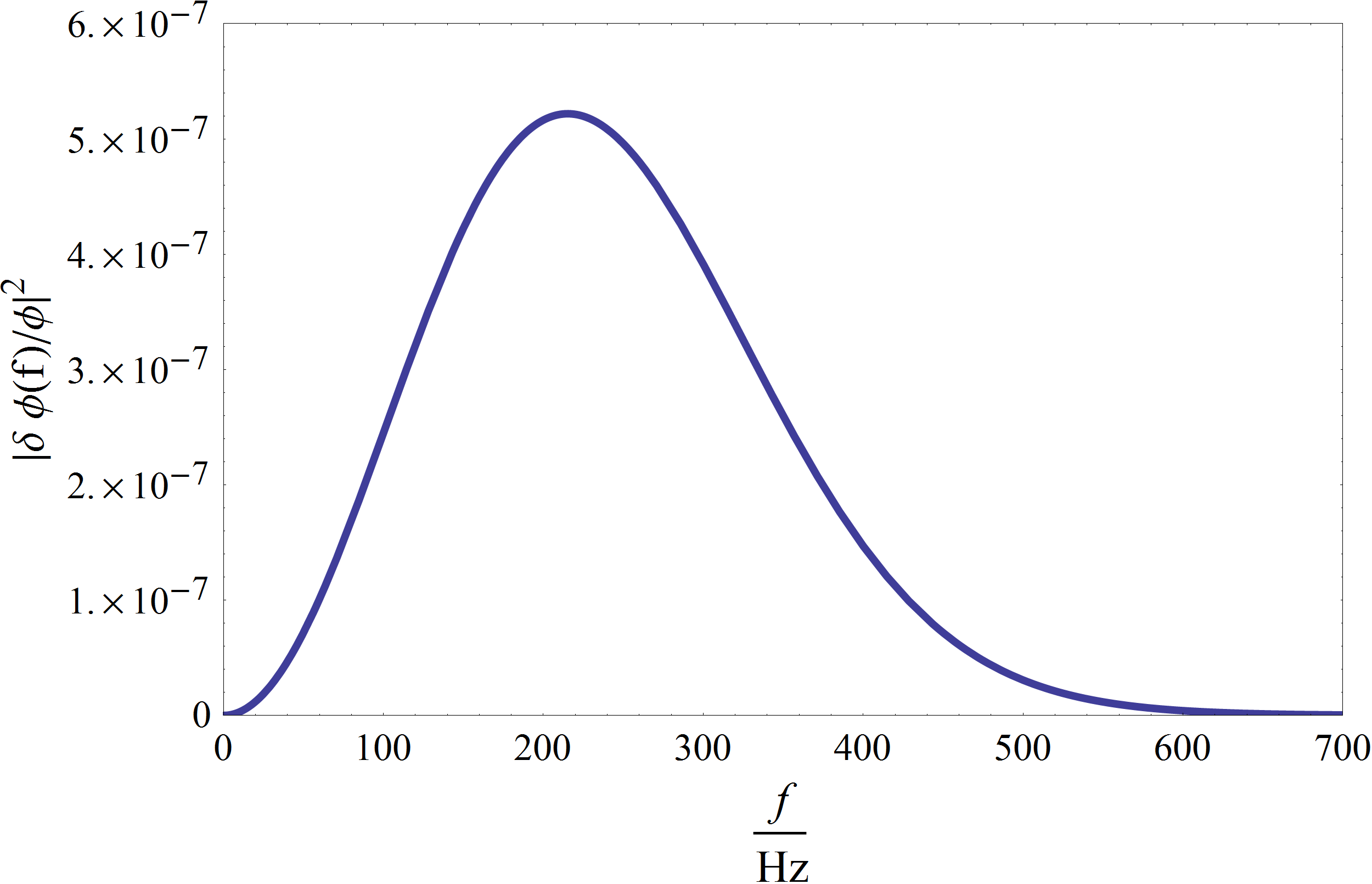}
\caption{(Color online) Power spectra [in units $ \rho_{\textrm{TDM}}^2 \tau^2 v^2 d^2 \hbar^2 c^2 / (\Lambda_X')^4 \times $s$^2$] versus frequency, produced by a domain wall with a Gaussian cross-sectional profile passing directly along one of the arms of a TAMA300 interferometer ($L=300$ m). From left to right: $d=1$ m, $d=10$ m, $d=100$ m, $d=300$ m.} 
\label{fig:TAMA}
\end{center}
\end{figure*}

\begin{figure*}[h!]
\begin{center}
\includegraphics[width=7cm]{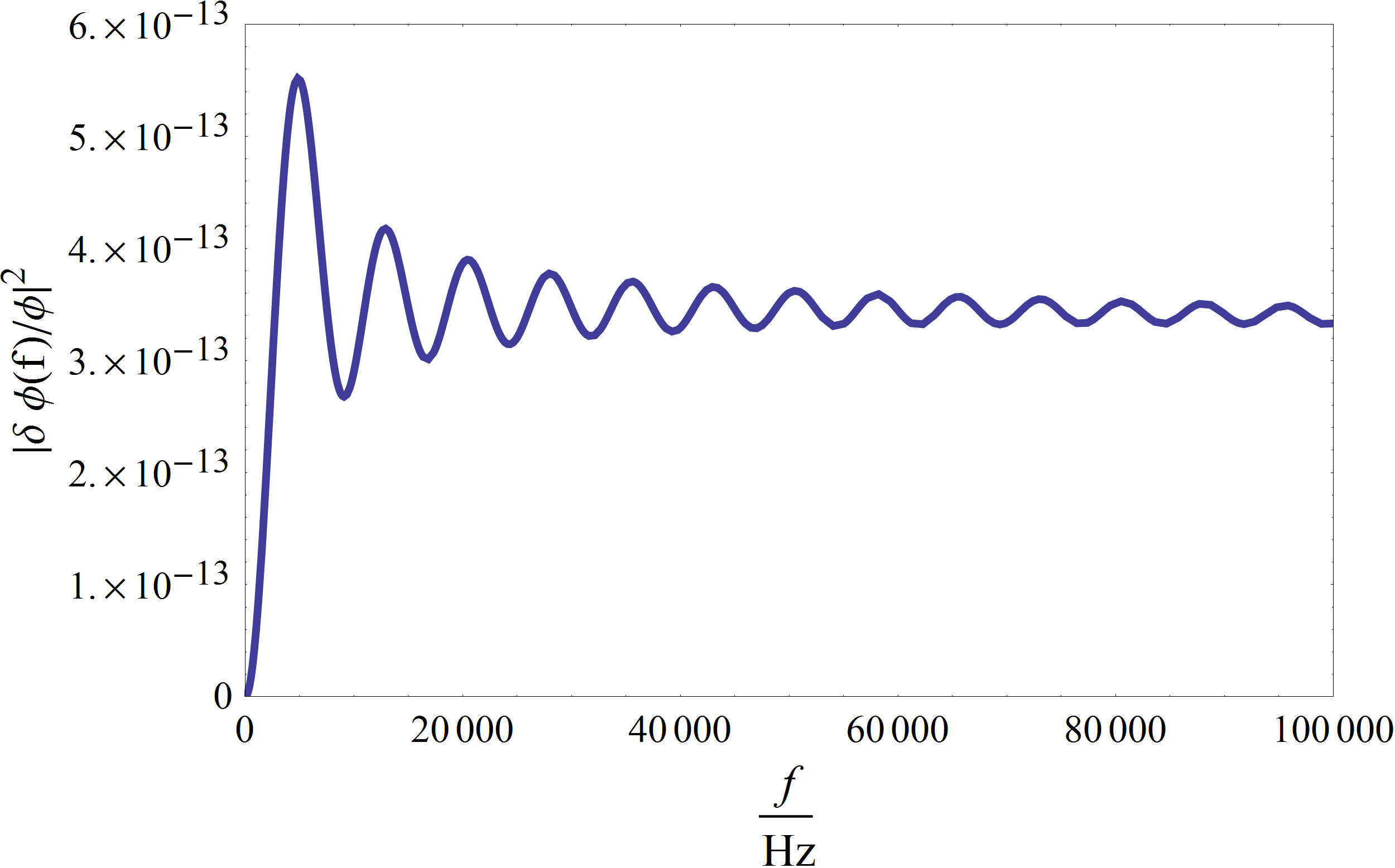}
\includegraphics[width=7cm]{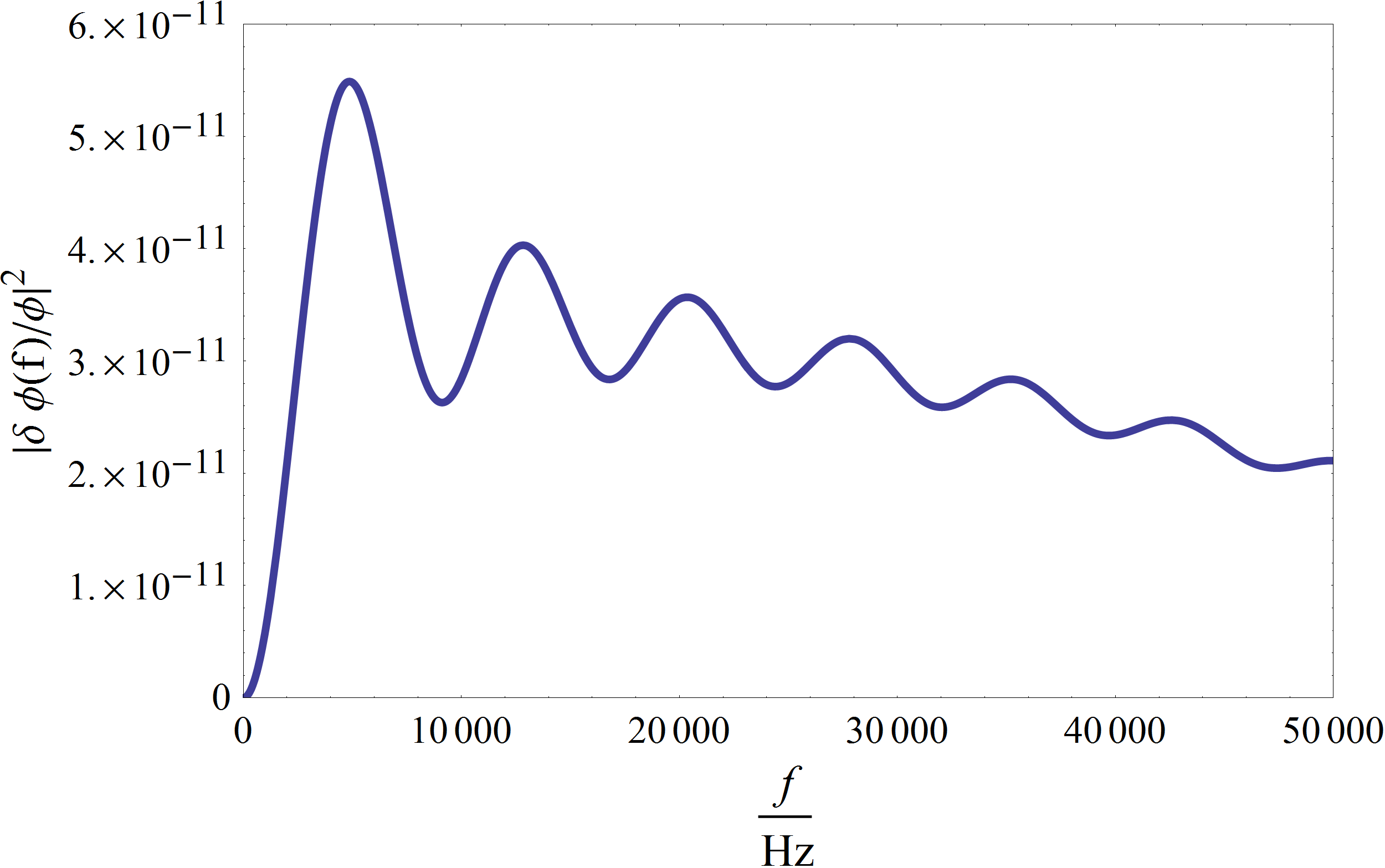}
\includegraphics[width=7cm]{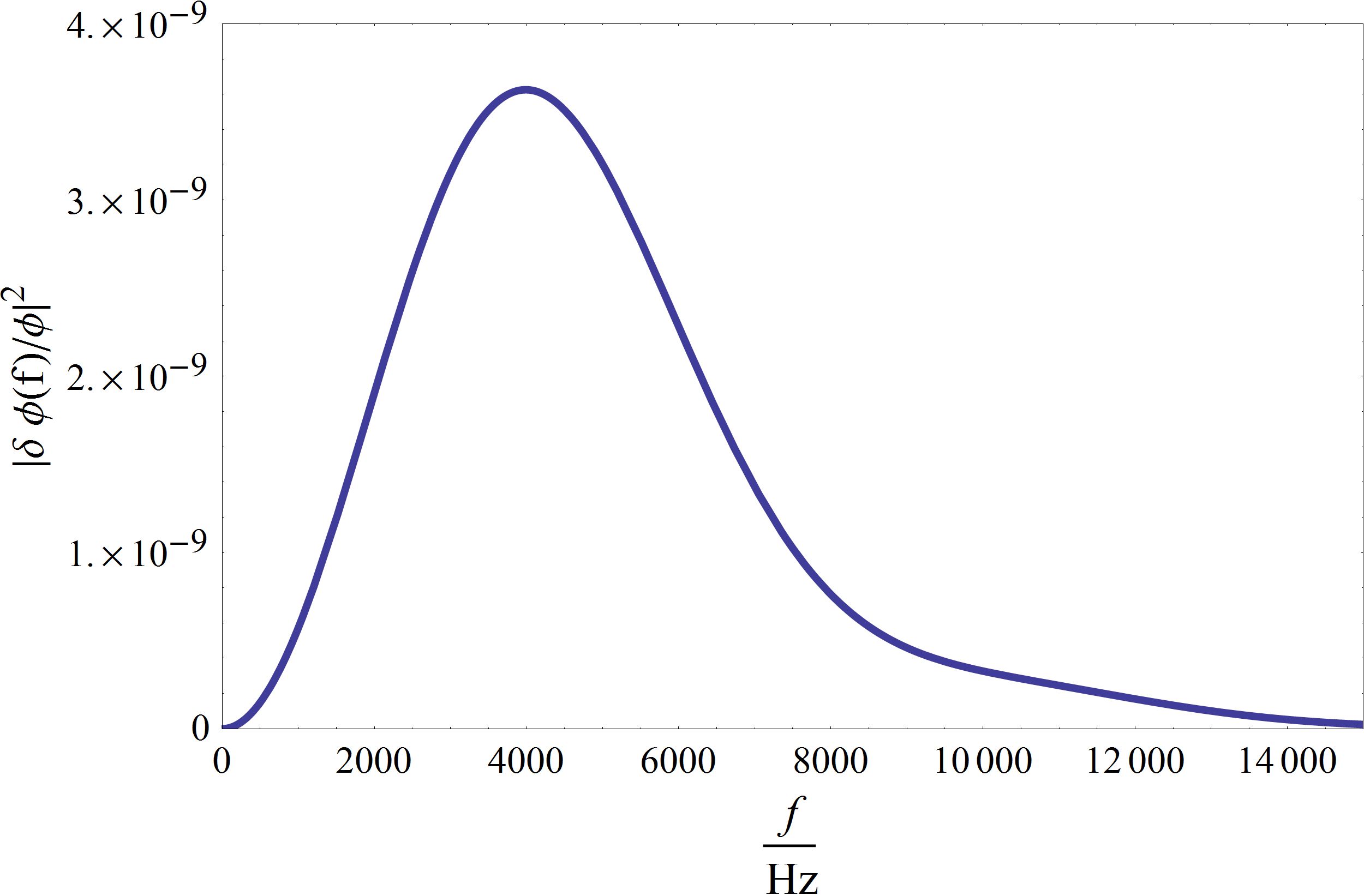}
\includegraphics[width=7cm]{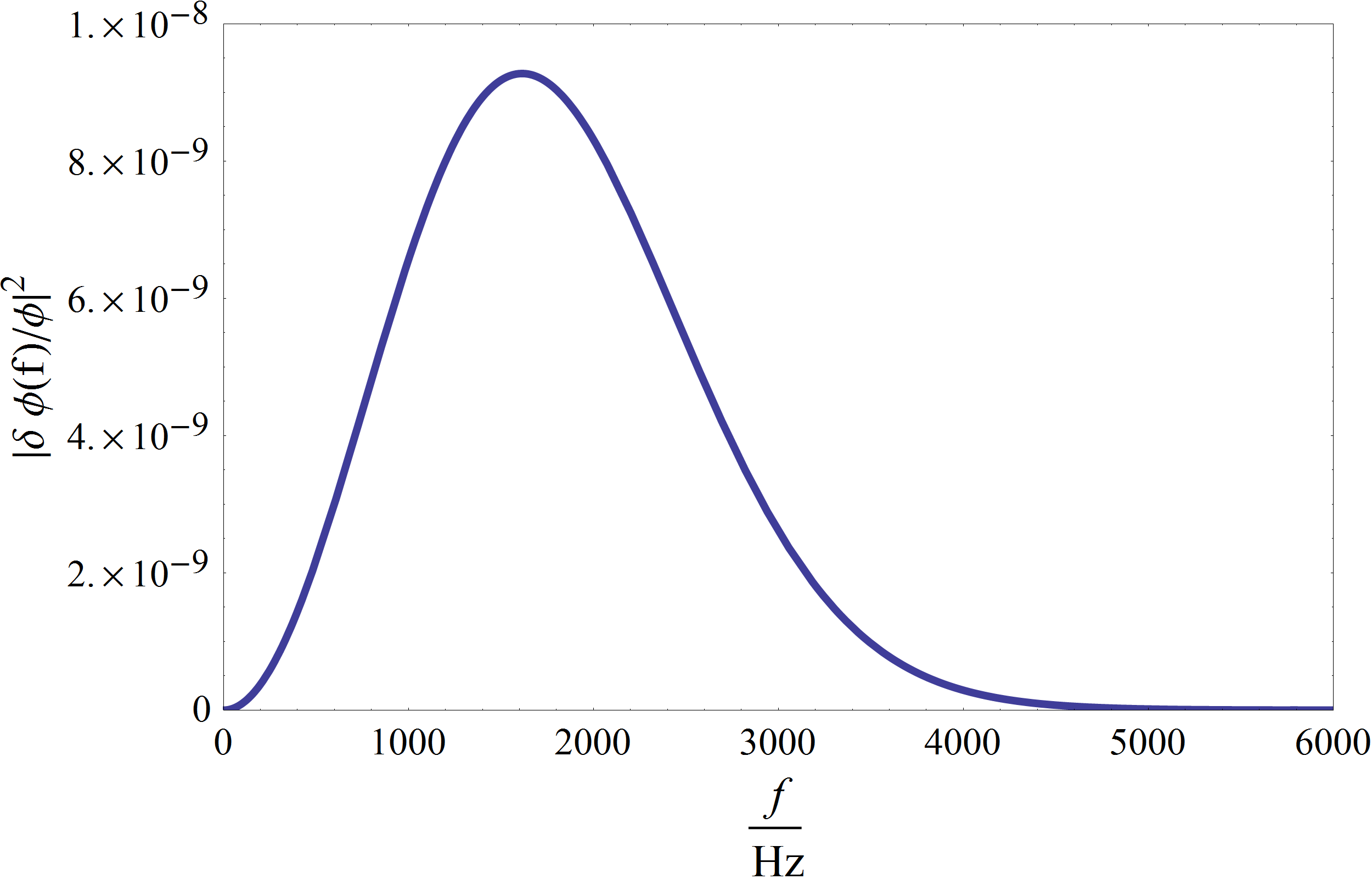}
\caption{(Color online) Power spectra [in units $ \rho_{\textrm{TDM}}^2 \tau^2 v^2 d^2 \hbar^2 c^2 / (\Lambda_X')^4 \times $s$^2$] versus frequency, produced by a domain wall with a Gaussian cross-sectional profile passing directly along one of the arms of a Fermilab Holometer interferometer ($L=40$ m). From left to right: $d=0.1$ m, $d=1$ m, $d=10$ m, $d=40$ m.} 
\label{fig:Fermilab}
\end{center}
\end{figure*}

We present the power spectra produced by a domain wall with a Gaussian cross-sectional profile of root-mean-square width $d$, passing directly along one of the arms of a Michelson-Morley interferometer. The asymptotic case $d \gg L$ can be summarised by a single plot in the combined variable $fd$ (Fig.~\ref{fig:d_larger_cf_L}). The plots in the general case $d \lesssim L$ are shown for the LIGO, Virgo, GEO600, TAMA300 and Fermilab Holometer interferometers in Figs.~\ref{fig:LIGO}, \ref{fig:Virgo}, \ref{fig:GEO}, \ref{fig:TAMA} and \ref{fig:Fermilab}, respectively.

\end{document}